\documentclass[useAMS,usenatbib,usegraphicx]{mn2e}
\usepackage{latexsym,graphicx,natbib,amssymb}

\usepackage{color}

\newcommand\kms{{\rm\,km\,s^{-1}}} 
\newcommand\kpc{{\rm\,kpc}} 
 
\newcommand\msun{\rm\,M_\odot}
\newcommand\lsun{\rm\,L_\odot}

\title[Velocity dispersion and mass function of Pal\,4]
 {The velocity dispersion and mass function of the outer halo globular cluster
Palomar~4}
      
\author[M. J. Frank et al.]
{Matthias J. Frank$^{1}$\thanks{E-mail: mfrank@ari.uni-heidelberg.de}, Michael
Hilker$^{2}$, Holger Baumgardt$^{3}$, Patrick
C\^ot\'e$^{4}$, \newauthor Eva K. Grebel$^{1}$, Hosein Haghi$^{5}$, Andreas H.
W. K\"upper$^{6}$, and S. G. Djorgovski$^{7,8}$\\
 $^1$ Astronomisches Rechen-Institut, Zentrum f\"ur Astronomie der
Universit\"at
Heidelberg, M\"onchhofstrasse 12~-~14, D-69120 Heidelberg, Germany \\
 $^2$ European Southern Observatory, D-85748 Garching b. M\"unchen, Germany \\
 $^3$ School of Mathematics and Physics, The University of Queensland,
Brisbane,
QLD 4072, Australia \\ 
 $^4$ Herzberg Institute of Astrophysics, National Research Council of Canada,
Victoria, BC V9E 2E7, Canada \\
 $^5$ Institute for Advanced Studies in Basic Sciences (IASBS), P.O.Box
45195-1159, Zanjan 4513766731, Iran \\
 $^6$ Argelander-Institut f\"ur Astronomie, Auf dem H\"ugel 71, D-53121 Bonn,
Germany \\
 $^7$ Astronomy Department, California Institute of Technology, Pasadena,
CA~91125, USA\\
 $^8$ Distinguished Visiting Professor, King Abdulaziz University, 21589 Jeddah,
Saudi Arabia
}

\date{Accepted 2012 April 13. Received 2012 April 13; in original form 2011 November 30}

\pagerange{\pageref{firstpage}--\pageref{lastpage}} \pubyear{2011}
\begin{document}   

\maketitle

\label{firstpage}

\begin{abstract}
We obtained precise line-of-sight radial velocities of 23 member stars of the
remote halo globular cluster Palomar~4 (Pal\,4) using the High Resolution
Echelle Spectrograph (HIRES) at the Keck I telescope. We also measured the
mass
function of the cluster down to a limiting magnitude of $V\sim28{\rm\,mag}$
using archival \textit{HST}/WFPC2 imaging. We derived the cluster's surface brightness 
profile based on the WFPC2 data and on broad-band imaging with the Low-Resolution
Imaging Spectrometer 
(LRIS) at the Keck II telescope.
We find a mean cluster velocity of $72.55 \pm 0.22\kms$ and a velocity
dispersion of $0.87 \pm 0.18\kms$. The global mass function of the cluster, in
the
mass range $0.55\le \mathrm{M} \le 0.85\msun$, is shallower than a
Kroupa mass function and the cluster is significantly depleted in low-mass
stars in its center compared to its outskirts. Since the relaxation time of
Pal\,4 is of the order of a Hubble time, this points to primordial mass
segregation in this cluster. Extrapolating the measured mass function towards
lower-mass stars and including the contribution of compact remnants, we derive
a
total cluster mass of 29,800\,M$_\odot$. For this mass, the measured velocity
dispersion is consistent with the expectations of Newtonian dynamics and below
the prediction of MOND. Pal\,4 adds to the growing body of evidence that the
dynamics of star clusters in the outer Galactic halo can hardly be explained
by MOND.
\end{abstract}


\section{Introduction}
\label{sec:intro}

The globular cluster (GC) system of the Milky Way extends out to more than
100\,kpc. Due to their old age and robust nature, GCs are believed to be
important tracers of the formation and early evolution of the Galaxy and its
halo. Of the more than 150 Galactic GCs \citep[e.g.][]{1996AJ....112.1487H}, 
about one quarter belongs to the so-called `outer halo', at Galactocentric 
distances larger than 15\,kpc \citep[e.g.][]{2004MNRAS.354..713V}. Most of
these are also attributed to the `young halo' GC sub-population because 
they seem to be 1-2~Gyr younger than the old, inner halo GCs of similar 
metallicity \citep[e.g.][]{2010ApJ...708..698D}. A number of authors have
suggested 
that the young and/or outer halo GCs were accreted by the Milky Way 
via the infall of dwarf satellite galaxies
\citep[e.g.][]{1996ASPC...92..434M,2000ApJ...533..869C,2004MNRAS.355..504M,
2007ApJ...661L..49L,2010MNRAS.404.1203F},
similar to the halo assembly scenario already proposed by
\citet{1978ApJ...225..357S}, 
whereas the old, inner GCs probably formed during an early and 
rapid dissipative collapse of the Galaxy's halo \`a la
\citet{1962ApJ...136..748E}.

Apart from being witnesses of the assembly of the Galactic halo, GCs are also 
valuable probes for testing fundamental physics
\citep[e.g.][]{2003A&A...405L..15S}. 
\citet{2005MNRAS.359L...1B} proposed to use diffuse outer halo GCs to 
distinguish between classical and modified Newtonian dynamics
\citep[MOND;][]{1983ApJ...270..371M,1983ApJ...270..365M,1984ApJ...286....7B}. 
MOND is very successful in explaining the flat rotation curves of disk
galaxies, without any assumption of unseen dark matter. According to MOND,
Newtonian 
dynamics breaks down for accelerations lower than $a_0 \simeq 1\times10^{-8}$
cm\,s$^{-2}$ \citep{1991MNRAS.249..523B,2002ARA&A..40..263S}. The external
acceleration due to the Galaxy experienced by remote outer halo clusters is
below this critical limit of $a_0$, and the radial velocity dispersion
profiles of such clusters can thus be used to distinguish between MOND and
Newtonian dynamics. \citet{2003A&A...405L..15S,
2007A&A...462L...9S}, \citet{2010A&A...523A..43S} and
\citet{2011A&A...525A.148S} reported a flattening of the velocity dispersion profile at accelerations
comparable to $a_0$ also in GCs with Galactocentric distances $\la20$~kpc.
However, as the external acceleration in these clusters is well above $a_0$,
such flattened velocity dispersion profiles in `inner' GCs are more commonly
attributed to the effects of tidal heating and unbound stars or to
contamination by field stars
\citep[e.g.][]{1998AJ....115..708D,2010MNRAS.401.2521L,2010MNRAS.406.2732L,2010MNRAS.407.2241K}.

In the context of testing MOND the massive outer halo cluster NGC~2419 has
received recent attention: Based on radial velocities of 40 of its members and
assuming isotropic stellar orbits, \citet{2009MNRAS.396.2051B} derived a
dynamical mass of 9$\pm2\times10^{5}\msun$, compatible with the photometric
expectation from a simple stellar population with a
\citet{2001MNRAS.322..231K} IMF. Moreover, they found no flattening of the
velocity dispersion profile at low accelerations that could point to MONDian
dynamics or dark matter in this cluster. \citet{2011ApJ...738..186I} studied
an extended radial velocity sample of 178 stars of NGC~2419 and found that,
while radial anisotropy is required in both Newtonian and MONDian dynamics to
explain the observed kinematics, the data favor Newtonian dynamics, with
their best-fitting MONDian model being less likely by a factor of $\sim40,000$
than their best-fitting Newtonian model. \citet{2012MNRAS.419L...6S}
challenged this conclusion, arguing that in MONDian dynamics non-isothermal
models, approximated by high-order polytropic spheres, can reproduce the
cluster's surface brightness and velocity dispersion profiles. This led
\citet{2011ApJ...743...43I} to extend the analysis of their data to polytropic
models in MOND. Again, they concluded that the best-fitting MONDian model is
less likely by a factor of $\sim5000$ than the best-fitting Newtonian model,
and that the data therefore pose a challenge to MOND, unless systematics are
present in the data \citep[but see also][]{2012MNRAS.422L..21S}. 

In the most diffuse outer halo clusters, i.e. clusters with large effective
radii, 
low masses and therefore low stellar densities, also the internal
acceleration 
due to the cluster stars themselves is below $a_0$ throughout the cluster. In
these clusters, not only the shape of the velocity dispersion profile, but
also the \emph{global} velocity dispersions can be used to discriminate
between MONDian and Newtonian dynamics.  \citet{2005MNRAS.359L...1B} showed
that the expected global velocity 
dispersions in case of MOND exceed those expected in the classical Newtonian 
framework by up to a factor of 3 in these clusters (see their table\,1). This
result was 
reinforced by more accurate numerical simulations including the external 
field effect by \citet{2011A&A...527A..33H}.

This paper continues a series of papers that investigates theoretically and
observationally the dynamics of distant, low-mass star clusters. In the first
paper \citep{2009MNRAS.395.1549H}, we derived theoretical models for
pressure-supported stellar systems in general and made predictions for the
outer halo GC Pal\,14 at a Galactocentric distance of about 72 kpc.
In the corresponding observational study of Pal\,14
\citep{2009AJ....137.4586J}, 
we showed that the observed velocity dispersion (based on 16 stars) and 
photometric mass of the cluster favor Newtonian dynamics over MOND. 

\citet{2010A&A...509A..97G} however, argued on the basis of a
Kolmogorov-Smirnov
test, 
that the sample of member stars in Pal\,14, (or, alternatively, the sample of
studied diffuse outer
halo GCs) 
is too small to rule out MOND. \citet{2010ApJ...716..776K}, reanalyzed the 
\citet{2009AJ....137.4586J} radial velocity data including a heuristic
treatment 
of binaries and mass segregation, and argued that Pal\,14 either has to 
have a very low binary fraction of less than 10 per cent or 
otherwise is in a `deep freeze' state, with an intrinsic velocity 
dispersion (after correction for binarity) low enough to challenge 
Newtonian dynamics in the opposite sense of MOND. However,
\citet{2012ApJ...744..196S} 
in a similar analysis of the same radial velocity data, found that the cluster
is compatible
with Newtonian dynamics also when the constraint of the binary fraction is
relaxed to $<30$ per cent.
Finally, the presence of tidal tails around Pal\,14
\citep{2010A&A...522A..71J,2011ApJ...726...47S} 
indicates that the cluster currently is undergoing tidal stripping, further
complicating the interpretation of its stellar kinematics.

In this paper, we present the internal velocity dispersion, the stellar mass
function and total stellar mass of the remote halo globular cluster Pal\,4.
With a Galactocentric
distance of 103\,kpc (see Section~\ref{sec:photresults:age}) it is the second
to
outermost halo GC after AM\,1 \citep[at 123\,kpc according to the 2010 edition
of the Galactic GC data base by][]{1996AJ....112.1487H}. Pal\,4 also is among
the most
extended Galactic GCs: its half-light radius of 18\,pc
(Section~\ref{sec:surfacebrightness}) is more than five times larger than that
of `typical' GCs \citep[e.g.][]{2005ApJ...634.1002J}. The cluster thus has a
size comparable to some of the Galaxy's ultra-faint dwarf spheroidal
satellites, but is at the same time brighter by $\sim2{\rm\,mag}$ in $V$ than
these \citep[e.g.][]{2007ApJ...654..897B}. 

Regarding its horizontal branch, Pal\,4 forms a so-called `second parameter
pair' with the equal-metallicity inner halo GC M\,5
\citep[e.g.][]{2000ApJ...531..826C}. Pal\,4 has a red horizontal branch and
M\,5
a blue one. One of the differences between M\,5 and Pal\,4 is their age.
Pal\,4
was found to be $\sim$1-2~Gyr younger ($\sim$10-11~Gyr) than M\,5
\citep{1999AJ....117..247S,2000ApJS..129..315V}. As mentioned above, such
relatively young halo clusters are thought to have been accreted from
disrupted
dwarf satellites. In this context, \citet{2010ApJ...718.1128L} discuss
Pal\,4's
possible association with the Sagittarius stream, but conclude that this is
unlikely based on current observational data and models of the stream's
location. In deep wide-field imaging of the cluster and its surroundings,
\citet{2003AJ....126..803S} find indications for the presence of extra-tidal
stars, but no significant detection of a stream. They attribute this
extra-tidal
overdensity to internal evaporation and tidal loss of stars at the cluster's
location in the Galaxy. 

The most recent determination of the chemical composition of Pal\,4 was
presented by
\citet{2010A&A...517A..59K}.  According to their abundance analysis of the
same
spectra that we use for our kinematical study, Pal\,4 has a metallicity of
[Fe/H]$=-1.41\pm0.17$\,dex and an $\alpha$-element enhancement of
[$\alpha$/Fe]$=0.38\pm0.11$\,dex. The metallicity is compatible with a
previous spectroscopic measurement of [Fe/H]$=-1.28\pm0.20$~dex by
\citet{1992AJ....104..164A}. 

This paper is organized as follows. In Section\,\ref{sec:obs}, we describe
the spectroscopic and photometric data and their reduction. In
Section\,\ref{sec:specresults}, we present stellar
radial velocities and the cluster's systemic velocity and velocity dispersion.
In
Section \ref{sec:photresults} we derive the cluster's surface brightness
profile, mass function and total stellar mass, and we present evidence
for mass segregation in the cluster. In Section\,\ref{sec:disc}, we discuss
our results with respect to expectations from classical Newtonian
gravity and MOND. The last section concludes the paper with a summary.

\section{Observations and Data Reduction}
\label{sec:obs}
Our analysis of the dynamical behavior of Pal\,4 is based on spectroscopic and
photometric observations. The High Resolution Echelle Spectrograph (HIRES) on
the Keck I telescope was used to obtain radial velocities and to derive the
velocity dispersion of Pal\,4's probable member stars. Pre-images for the
spectroscopy were obtained with the Low-Resolution Imaging Spectrometer (LRIS)
mounted on the Keck II telescope and used to derive the cluster's structural
parameters. Both Keck datasets are part of a larger program dedicated to study
the internal kinematics of outer halo GCs \citep[for details of the program
see][]{2002ApJ...574..783C}. Archival imaging data obtained with the Hubble
Space Telescope's (\textit{HST}) Wide Field Planetary Camera 2 (WFPC2) were analyzed to
determine the mass function and total mass of the cluster.

\begin{figure}
\includegraphics[width=84mm]{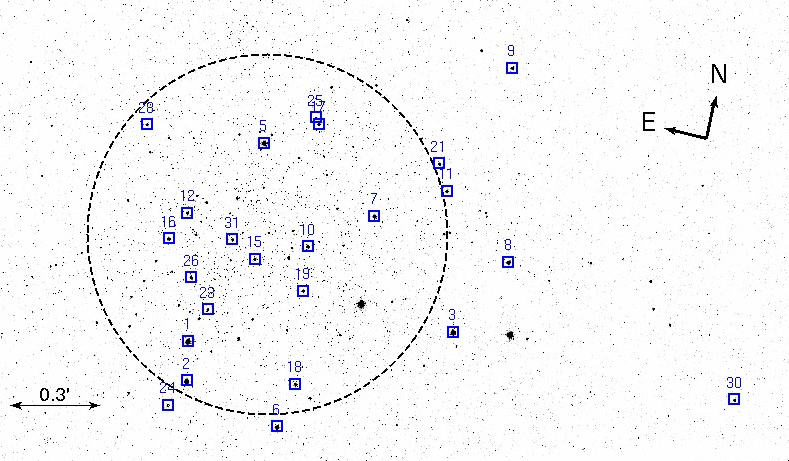}
\caption{Position of spectroscopic target stars on the sky, overlaid on an
archival \textit{HST} Advanced Camera for Surveys (ACS) image (program 10622, PI:
Dolphin). The numbering corresponds to the order of objects as listed in
Table\,\ref{tab:radvel}. The dotted circle marks Pal\,4's half-light radius of
$0.6$ arcmin, corresponding to 18 pc at a distance of 102.8 kpc.}
\label{fig:pal4finder}
\end{figure}

\subsection{Keck LRIS Photometry}
$B$ and $V$ images centered on Pal\,4 were obtained with LRIS
\citep{1995PASP..107..375O} on the night of 1999 January 14. In imaging mode,
LRIS has a pixel scale of 0.215\,arcsec\,pixel$^{-1}$ and a field of view of
$5.8\times7.3$ arcmin$^2$. A series of images were obtained in  both $V$ and $B$,
with exposure times of $3\times60$s and $2\times180$s, respectively.
Conditions during the night were photometric, and the FWHM of isolated stars
within the frames was measured to be 0\farcs65--0\farcs75. The images were
reduced in a manner identical to that described in \citet{2002ApJ...574..783C}
using IRAF\footnote{IRAF is distributed by the National Optical Astronomy
Observatories, which are operated by the Association of Universities for
Research in Astronomy, Inc., under cooperative agreement with the National
Science Foundation.}. Briefly, the raw frames were bias-subtracted and
flat-fielded using sky flats obtained during twilight. Instrumental magnitudes
for unresolved objects in the field were derived using the DAOPHOT II software
package \citep{1993spct.conf..291S}, and calibrated with observations of
several \citet{1992AJ....104..340L} standard fields taken throughout the
night. The $V$ band magnitudes, which we used to calibrate the cluster's
surface brightness profile (Section~\ref{sec:surfacebrightness}), were found
to agree to within 0.02$\pm$0.03~mag with those published by
\citet{2005PASP..117...37S} for stars contained in both catalogs. The final
photometric catalog contained 848 objects detected with a minimum point-source
signal-to-noise ratio of S/N = 4 in both filters. 

\subsection{Spectroscopy}
On three different nights in February and March 1999, spectra for 24 candidate
red giants in the direction of Pal\,4 were obtained using HIRES
\citep{1994SPIE.2198..362V} mounted on the
Keck\,I telescope. The targets were selected from the LRIS photometric
catalog. The spectra were taken with the C1 decker, which gives a $0\farcs86$
entrance slit and a resolution of $R=45,000$, and cover the wavelength range
from 445 to 688\,nm. 
Their position within the cluster is shown in Fig.\,\ref{fig:pal4finder}.
The exposure times of the spectra were adjusted on
a star-to-star basis depending on the individual magnitudes
($17.8<V<19.9{\rm\,mag}$), and varied between 300 and 2400\,s with a median
value of 1200\,s. An observation log and the photometric properties of the
target stars are given in Table~\ref{tab:radvel}, their coordinates are given
in table\,1 of \citet{2010A&A...517A..59K}, and their location in the
color-magnitude diagram (CMD) can be seen in fig.\,1 of the same paper.
Based on their location in the CMD, five of the sample stars are probable AGB
stars, the remaining 19 stars lie on the RGB.

The spectra were reduced entirely within the IRAF environment, in a manner
identical to that described in
\citet{2002ApJ...574..783C}. The radial velocities of the target stars were
obtained by
cross-correlating their spectra with those of master templates created from
the
observations of IAU standard stars, which were taken during the seven
observing runs (13 nights) that were devoted to the HIRES survey of globular
clusters in the halo. From each cross-correlation function, we measured the
heliocentric
radial velocity, $v_r$, and $R_{\rm TD}$, the \citet{1979AJ.....84.1511T}
estimator of the strength of the cross-correlation peak. Since an important
factor in the dynamical analysis of low-mass clusters is an accurate
determination of the radial velocity uncertainties, $\epsilon(v_r)$, 53 repeat
measurements for 23 different stars, distributed over different target GCs,
were accumulated during the same observing runs. The r.m.s. of the repeat
measurements was used to calibrate a relation between $\epsilon(v_r)$ and
$R_{\rm TD}$. Following \citet{1995AJ....109..151V}, we adopt a relationship
of the form $\epsilon(v_r) = {\alpha}/(1 + R_{\rm TD})$, where $R_{\rm TD}$ is
the \citet{1979AJ.....84.1511T} estimator of the strength of the
cross-correlation peak, and find $\alpha \simeq 9.0$ km~s$^{-1}$. The
resulting radial velocity uncertainties for our Pal\,4 target stars range from
$0.23$ to $1.31\,\kms$ (see Table~\ref{tab:radvel}).

\subsection{\textit{HST} Photometry}
We used archival \textit{HST} images of Pal\,4 obtained with the WFPC2 in GO program
5672 \citep[PI: Hesser, cf.][]{1999AJ....117..247S}. The dataset consists of
F555W
($V$) and F814W ($I$) band exposures and is the deepest
available broad-band imaging of the cluster. The individual exposure times are
8$\times$30\,s, 8$\times$60\,s and
8$\times\sim1800$\,s in each filter, amounting to total exposure time of
$\sim4.1$\,h per filter.

PSF-fitting photometry was obtained using the \textsc{HSTPHOT} package
\citep{2000PASP..112.1383D}. In order to refine the image registration,
\textsc{HSTPHOT}
was first run on the individual images and the resulting catalogs were matched
to
one of the deep F555W images as a reference using the IRAF tasks xyxymatch and
geomap. The derived residual shifts were used for a refined cosmic ray
rejection
with \textsc{HSTPHOT}'s crmask task, and as an input for the photometry from all
images.
The latter was obtained by running \textsc{HSTPHOT} simultaneously on all frames with a
deep F555W image as the reference or detection image. 

\begin{figure}
\includegraphics[width=84mm]{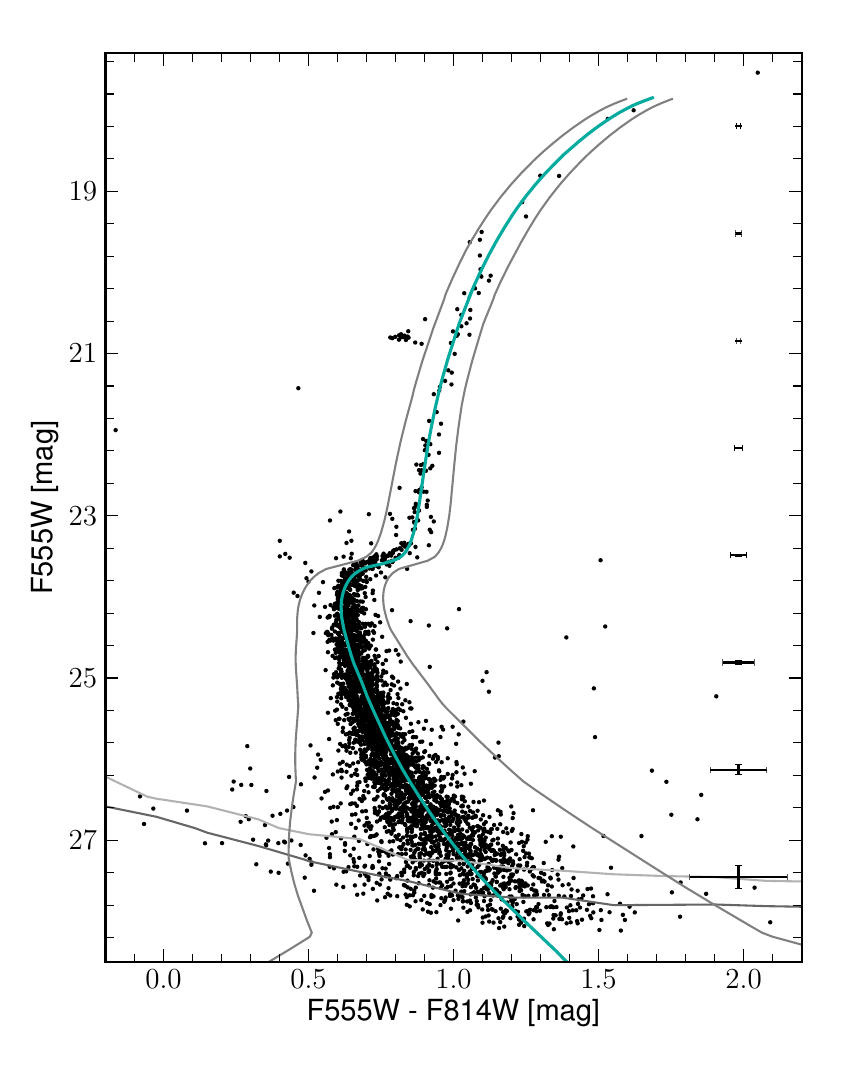}
\caption{Observed color-magnitude diagram of Pal\,4. Errorbars on the right
represent the photometric errors derived from artificial star tests. The gray
lines at the faint end represent the 80 per cent (light gray) and 50 per cent (dark gray)
completeness contours. The isochrone (cyan line) corresponds to an age of
11~Gyr, a
metallicity of [Fe/H]=$-1.41$~dex and an $\alpha$-enhancement of
[$\alpha$/Fe]=$+0.4$~dex, shifted to the cluster's distance of
$20.06{\rm\,mag}$
at a reddening of $\mathrm{E}(B$-$V)=0.023{\rm\,mag}$. Thin gray curves to the
left
and to the right of the isochrone represent the  color limits used for our
analysis of the cluster's mass function
(see Section~\ref{sec:photresults}).}
\label{fig:CMD}
\end{figure}

To select bona-fide stars from the output catalog, the following quality cuts
were applied (for details, see the \textsc{HSTPHOT} user manual): a type parameter of 1
(i.e. a stellar detection), abs(sharpness)$<0.2$, $\chi<2.0$, and in both
filters a crowding parameter $<1.5{\rm\,mag}$ and a statistical uncertainty in
the magnitude $<0.2{\rm\,mag}$. The resulting CMD, containing 3878 stars, is
shown
in Fig.~\ref{fig:CMD}.
\begin{figure}
\includegraphics[width=84mm]{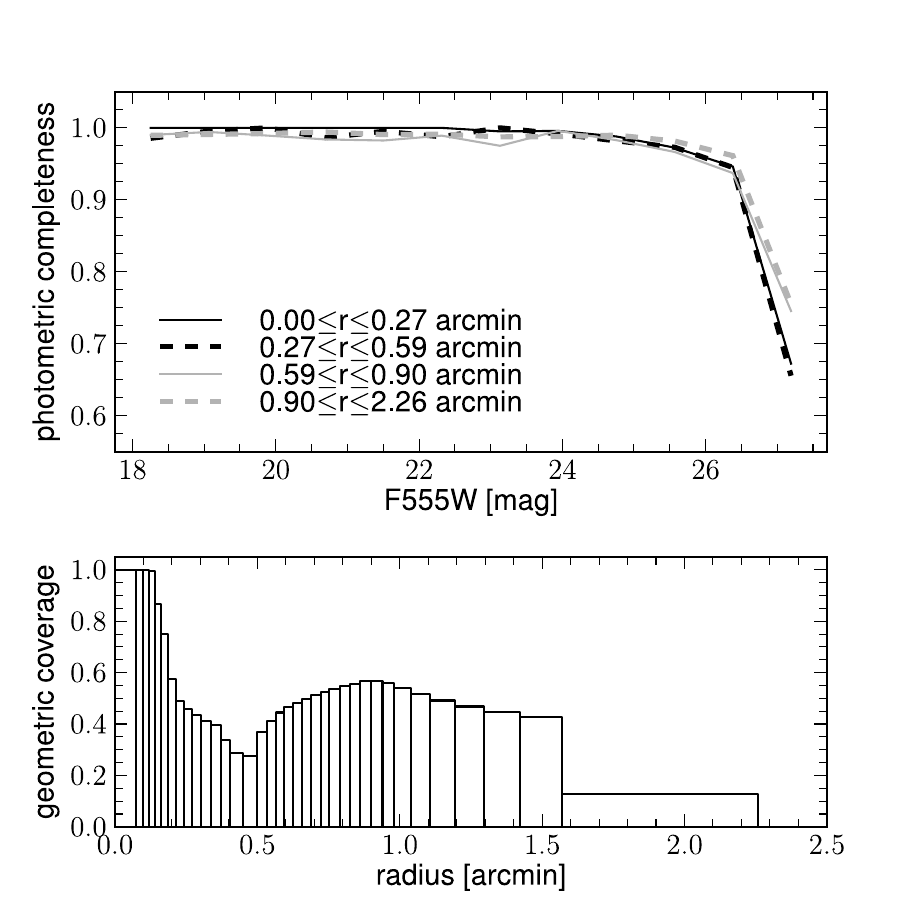}
\caption{Photometric and geometric completeness of the WFPC2 photometry. Top:
The photometric completeness inside the color limits used for our analysis
(see Fig.~\ref{fig:CMD}) as a function of F555W magnitude derived from the
artificial star tests is shown for four radial ranges as denoted in the plot.
The radial ranges are defined to contain one fourth of the observed stars
each. Bottom:
The geometric coverage of the WFPC2 catalog as a function of radius in radial
bins containing each one 36th of the observed stars. The fraction represents
the area covered by the WFPC2 pointing in a given radial annulus divided by
the total area of the annulus.}
\label{fig:completeness}
\end{figure}
To assess the photometric uncertainties and completeness of the catalog,
\textsc{HSTPHOT}
was used to perform artificial star tests with $\sim 275000$ fake stars. We
used
the program's option to create artificial stars with distributions similar to
the observed stars, both in the CMD, and on the WFPC2 chips, in order to
efficiently sample the relevant parameter space. In artificial star mode, the
program inserts, star by star, stellar images with given magnitudes and
position
in all of the frames (using the empirically adjusted PSF for each frame that
is
constructed during the photometry run) and then performs photometry on this
stellar image. It yields as a result a catalog containing the inserted
magnitudes and positions, as well as the recovered photometry for each fake
star. We applied the same quality cuts to the artificial star catalog as were
used to select bona-fide stars in the observed catalog. Photometric
uncertainties in a given region of the CMD and on the sky were then estimated
from the differences between inserted and recovered magnitudes. The
photometric
completeness was estimated from the ratio of the number of recovered to the
number of inserted artificial stars. The completeness, within the color limits
used for our analysis of the cluster's mass function (see
Section~\ref{sec:photresults:massfunction}), as a function of F555W magnitude
is
shown in the top panel of Fig.~\ref{fig:completeness}. The different curves
correspond to the completeness in different radial ranges, containing each one
fourth of the observed stars. At the faint end, the completeness in the inner
two annuli drops somewhat faster with decreasing luminosity, which reflects
the
effect of crowding caused by the higher surface density of stars in the
cluster's center.

The geometric coverage of the WFPC2 photometry was quantified in the following
way. For both filters, we ran multidrizzle \citep{2006hstc.conf..423K} on all
frames in that filter, to obtain geometric distortion-corrected combined
frames.
As a small-scale dither pattern was used in the observations, we then created
a
coverage mask by selecting all pixels that received, in both filters, at least
25 per cent of the total exposure time. This information can be retrieved from the
weight map extension of the drizzled frames. As \textsc{HSTPHOT} uses a single deep
exposure
as a detection image for the photometry, we additionally required that pixels
flagged as covered in the coverage mask were covered also by one of the four
chips in that exposure. For this, in order to avoid possible completeness
artifacts near chip borders,
the chips were assumed to be smaller by 5 pixels on each side. The area
covered
by the WFPC2 photometry as a function of distance from the cluster's center
was
then expressed as the ratio of the area covered by the coverage mask to the
total area of a given radial annulus around the cluster's center. This is
shown
in the bottom panel of Fig.~\ref{fig:completeness}. 
The stellar positions in the photometric and artificial star catalogs were 
transformed to the same drizzled coordinate system and to be consistent, stars
falling on pixels marked as `not covered' in the coverage mask were rejected.
In order to select radial subsamples of stars, we determined the cluster's
center by fitting one-dimensional Gaussians to the distributions of stars
projected onto the $x$ and $y$ axes \citep[e.g.][]{2006A&A...448..171H}. As
the cluster's center is close the PC chip's border in the WFPC2 pointing, for
the purpose of determining the center, we performed photometry on more
suitable archival data taken with the Wide Field Channel (WFC) of \textit{HST}'s
Advanced Camera for Surveys (ACS) in GO program 10622 \citep[PI: Dolphin,
cf.][]{2011PASP..123..481S}. We used \textsc{HSTPHOT}'s successor \textsc{DOLPHOT} on the
program's F555W (two exposures of 125\,s each) and F814W ($2\times80$\,s)
exposures to obtain a photometric point source catalog, determined the center
form these data and transformed its coordinates to the coordinate system of
the WFPC2 catalog.

\begin{figure}
\includegraphics[width=84mm]{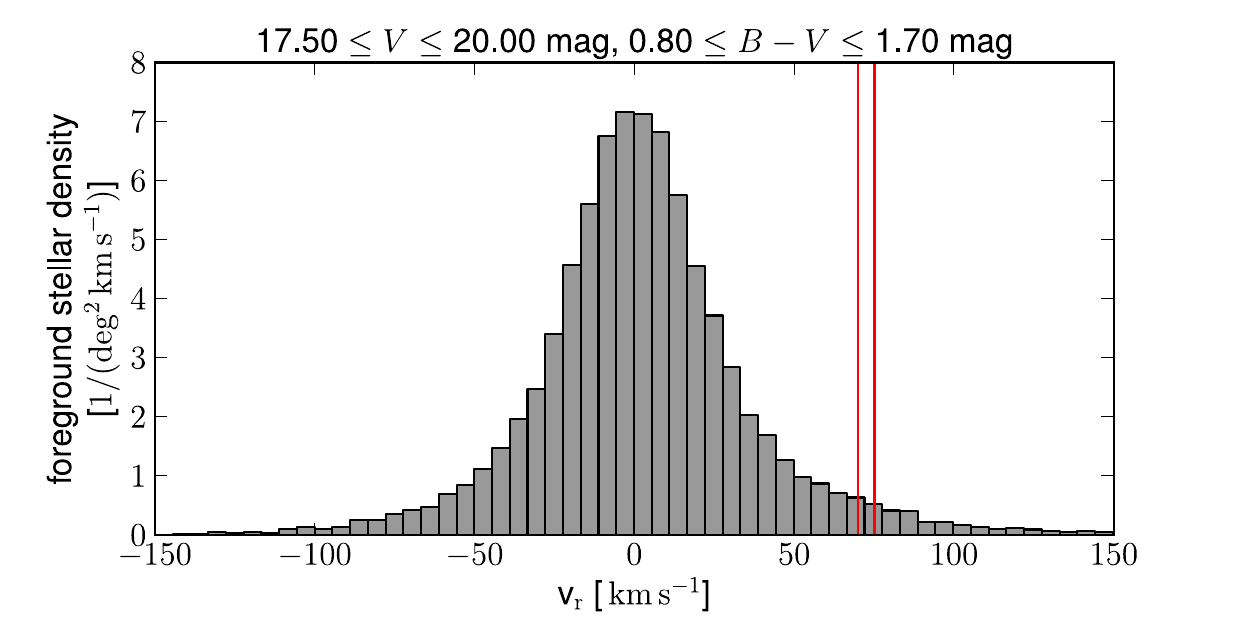}
\includegraphics[width=84mm]{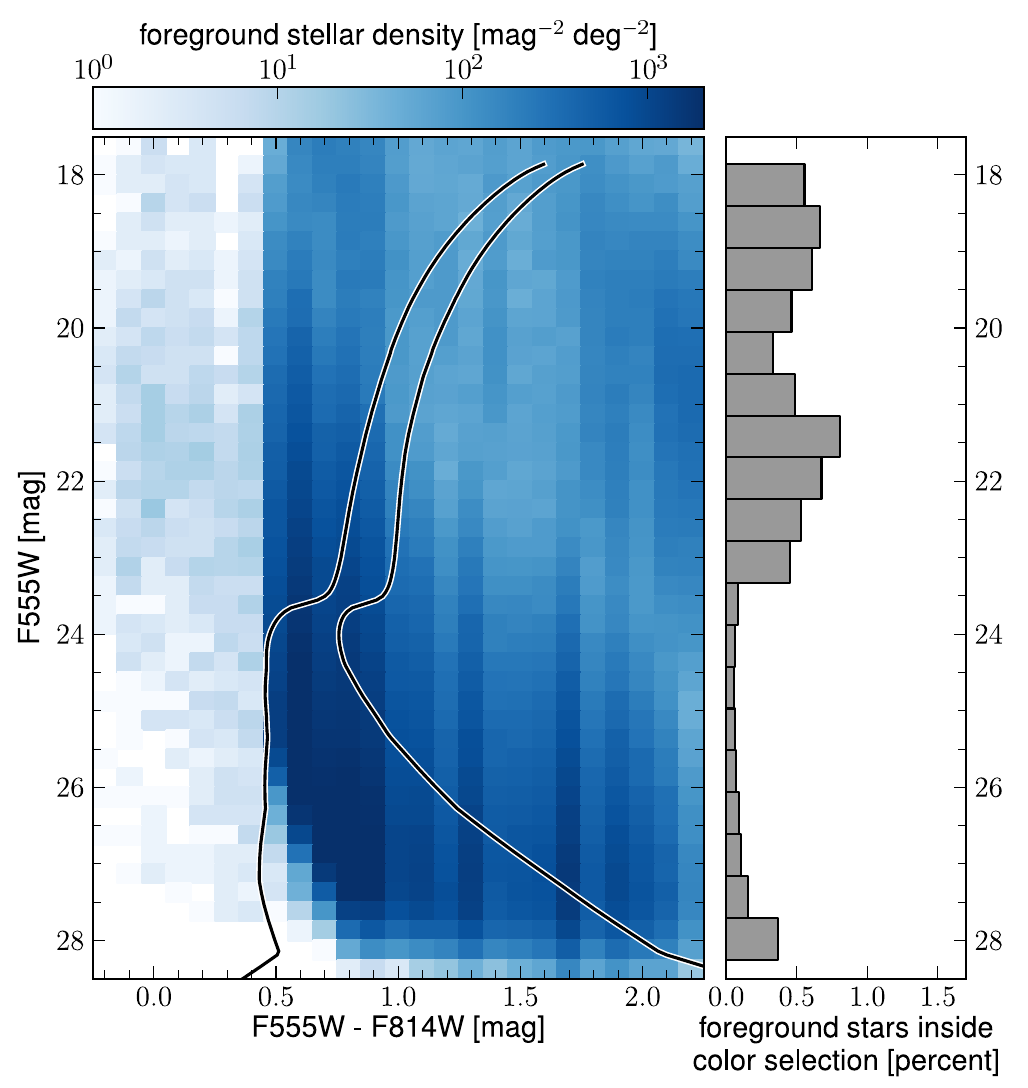}
\caption{Expected contamination by foreground stars based on the Besan\c{c}on
model. Top: The distribution of foreground stars having magnitudes and colors
in
the range of our spectroscopic targets as a function of radial velocity. Red
vertical lines denote the velocity range of interest. Bottom: The left panel
shows the density of foreground stars in the color-magnitude diagram. The
black-on-white lines
correspond to the region of the CMD used to estimate the mass function of
Pal\,4
(Section~\ref{sec:photresults:massfunction}). Within these color-limits, the
fraction of expected foreground stars in the photometric sample, averaged over 
$0.5$\,mag in F555W and shown in the right panel, is below one per cent.}
\label{fig:besancon}
\end{figure}

\subsection{Foreground contamination}
As Pal\,4 lies on ``our side'' of the Galaxy at high Galactic latitude
($l\,\sim$202\,deg, $b\,\sim$72\,deg), the expected contamination by
foreground
stars in our spectroscopic and photometric samples is low. To estimate its
fraction, we used the Besan\c{c}on model of the Galaxy
\citep{2003A&A...409..523R} to obtain a photometric and kinematic synthetic
catalog. The model was queried for stars out to
200$\kpc$ in the direction of Pal\,4. For better number statistics, we used 
a solid angle of 50 square degrees and the model's `small field' mode that
simulates
all stars at the same location and thus ensures that any spatial variation in
the foreground that could be
present in such a large field is neglected. The remaining model parameters,
such as the 
extinction law and spectral type coverage, were left at their default values.

For a generous estimate of possible foreground contaminants in our
spectroscopic
sample, we selected from the obtained synthetic catalog stars with magnitudes
and colors in the range of the spectroscopic targets (17.5 $\le V \le$ 20.0
mag,
0.8 $\le B-V \le$ 1.7 mag, cf. Table~\ref{tab:radvel}). The top panel of
Fig.~\ref{fig:besancon} shows the resulting distribution of stars per square
degree as a function of radial velocity. Red vertical lines denote the
velocity
range of the cluster's systemic velocity plus and minus three times its
velocity dispersion (derived in Section~\ref{sec:specresults}). Within this
velocity range, $\sim$3 stars per
deg$^{2}$ lie inside the color and magnitude range. Scaled to the solid angle
covered by the spectroscopic sample (assuming a circular aperture with a
radius
equal to the largest cluster-centric distance of our sample stars, $\sim
100$\,arcsec), this amounts to $\sim0.01$ stars. It is thus unlikely that the
spectroscopic sample contains any foreground stars.

To quantify the expected foreground contamination in the photometric catalog,
we
transformed the $V$ and $I$ magnitudes of the synthetic foreground stars to F555W and
F814W
magnitudes, by inverting the \citet{1995PASP..107.1065H} WFPC2 to $UBVRI$
transformations. Photometric errors and completeness were then taken into
account in the following simple way: for each synthetic foreground star, we
selected from our artificial star catalog a random one of the 100 nearest 
artificial stars in terms of inserted magnitudes (using the euclidean distance
in the (F555W, F814W)-plane); if the chosen artificial star was recovered, we
added its photometric errors (i.e. recovered minus inserted magnitude) to the
magnitudes of the synthetic star; if the artificial star was not recovered, we
reject the synthetic foreground star. To take into account the variation of
completeness and photometric errors as a function of distance from the cluster
center, we assumed the synthetic foreground stars to be homogeneously
distributed on the sky and performed the procedure independently on 30 radial
sub-samples of the foreground and artificial star catalogs. This results in a
foreground catalog that reproduces the photometric errors and completeness
limits of our WFPC2 catalog. The density
of foreground stars is shown in the bottom left panel of
Fig.~\ref{fig:besancon}. The two-dimensional histogram was obtained with
bins of 0.1~mag in color and 0.25~mag in magnitude and scaled to units of
stars per square degree on the sky and square magnitude in the CMD. Selecting
stars only in the region of the CMD that was
used to derive the mass function of Pal\,4 (denoted by the black-on-white
lines in the density
plot; see Section~\ref{sec:photresults:massfunction}) and scaling to the
effective area of the WFPC2 field, $\sim4.76$ arcmin$^2$, we calculated that
the expected
fraction of foreground stars in the photometric sample is below one per cent
over the whole
luminosity range and therefore negligible. This is shown in the bottom
right panel of Fig.~\ref{fig:besancon}.

\begin{table*}
\caption[]{Radial velocities for candidate red giants in Pal\,4. $(a)$Probable
AGB stars based on their location in the CMD.}
\begin{tabular}{llrccrrccc}
\hline\hline
 ID & ID$_{\rm Saha}$ & $R$ & $V$ & $(B-V)$ & $T$ & HJD 2,450,000+ &
 $R_{\rm TD}$ & $v_r$ & $\langle v_r\rangle$ \\
 & & (arcsec) & (mag) & (mag) & (s) & & & (km s$^{-1}$) & (km s$^{-1}$) \\
 (1) & (2) & (3) & (4) & (5) & (6) & (7) & (8) & (9) & (10) \\
\hline
Pal4-1  & S196 &  23.3 & 17.81 & 1.46 &  300 & 11220.9836 & 18.91 &
 73.59$\pm$0.45 & 73.33$\pm$0.28 \\
 & & & & &  300 & 11248.0317 & 16.50 & 72.84$\pm$0.52 & \\
 & & & & &  300 & 11221.1684 & 18.06 & 73.45$\pm$0.47 & \\
Pal4-2  & S169 &  29.9 & 17.93 & 1.46 &  300 & 11220.9787 & 16.61 &
 73.95$\pm$0.51 & 74.42$\pm$0.36 \\
 & & & & &  300 & 11221.1634 & 16.31 & 74.90$\pm$0.52 & \\
Pal4-3  & S277 &  41.2 & 17.82 & 1.66 &  300 & 11221.1388 & 20.09 &
 72.11$\pm$0.43 & 72.11$\pm$0.43 \\
Pal4-5  & S434 &  22.9 & 17.95 & 1.44 &  300 & 11221.1457 & 17.14 &
 72.24$\pm$0.50 & 72.41$\pm$0.41 \\
 & & & & &  300 & 11222.1754 & 11.36 & 72.78$\pm$0.73 & \\
Pal4-6  & S158 &  34.7 & 18.22 & 1.30 &  420 & 11220.9647 & 18.37 &
 72.34$\pm$0.47 & 72.38$\pm$0.33 \\
 & & & & &  420 & 11248.0018 & 10.42 & 72.47$\pm$0.79 & \\
 & & & & &  420 & 11221.1152 & 15.36 & 72.39$\pm$0.55 & \\
Pal4-7  & S381 &  23.6 & 18.55 & 1.19 &  600 & 11221.0986 & 17.44 &
 73.08$\pm$0.49 & 72.73$\pm$0.38 \\
 & & & & &  600 & 11248.0382 & 14.08 & 72.21$\pm$0.60 & \\
Pal4-8  & S364 &  49.4 & 18.65 & 1.17 &  600 & 11220.9989 & 16.59 &
 74.39$\pm$0.51 & 74.39$\pm$0.51 \\
Pal4-9  & S534 &  63.1 & 19.00 & 1.08 &  750 & 11221.0124 & 14.48 &
 71.56$\pm$0.58 & 71.56$\pm$0.58 \\
Pal4-10 & S325 &   8.9 & 19.09 & 1.05 &  900 & 11220.9880 & 16.83 &
 70.11$\pm$0.51 & 70.68$\pm$0.41 \\
 & & & & &  900 & 11221.1720 & 11.86 & 71.76$\pm$0.70 & \\
Pal4-11$^a$ & S430 &  39.2 & 19.35 & 0.89 & 1200 & 11221.0705 & 10.12 &
 73.08$\pm$0.81 & 73.08$\pm$0.81 \\
Pal4-12$^a$ & S328 &  18.1 & 19.35 & 0.90 & 1200 & 11221.1041 & 13.09 &
 78.70$\pm$0.64 & 76.22$\pm$0.43 \\
 & & & & & 1200 & 11247.9845 & 14.50 & 74.19$\pm$0.58 \\
Pal4-15$^a$ & S307 &   2.2 & 19.38 & 0.88 & 1200 & 11221.0550 &  9.62 &
 72.33$\pm$0.85 & 72.33$\pm$0.85 \\
Pal4-16$^a$ & S306 &  19.9 & 19.43 & 0.88 & 1200 & 11221.0383 & 13.05 &
 71.09$\pm$0.64 & 71.09$\pm$0.64 \\
Pal4-17$^a$ & S472 &  28.9 & 19.45 & 0.85 & 1080 & 11222.0903 & 11.67 &
 71.87$\pm$0.71 & 71.87$\pm$0.71 \\
Pal4-18 & S186 &  26.7 & 19.48 & 0.98 & 1200 & 11221.1275 & 12.23 &
 71.17$\pm$0.68 & 71.17$\pm$0.68 \\
Pal4-19 & S283 &  10.4 & 19.53 & 0.95 & 1080 & 11222.0760 & 10.41 &
 72.75$\pm$0.79 & 72.75$\pm$0.79 \\
Pal4-21 & S457 &  40.0 & 19.64 & 0.93 & 1200 & 11221.0869 &  9.53 &
 74.41$\pm$0.86 & 74.41$\pm$0.86 \\
Pal4-23 & S235 &  15.9 & 19.70 & 0.93 & 1500 & 11222.1575 & 12.43 &
 73.23$\pm$0.67 & 73.23$\pm$0.67 \\
Pal4-24 & S154 &  36.0 & 19.74 & 0.92 & 1500 & 11221.1612 & 13.50 &
 73.00$\pm$0.62 & 73.00$\pm$0.62 \\
Pal4-25 & S476 &  29.9 & 19.77 & 0.91 & 1500 & 11222.1782 &  9.41 &
 72.84$\pm$0.87 & 72.84$\pm$0.87 \\
Pal4-26 & S265 &  15.7 & 19.83 & 0.91 & 1500 & 11222.1389 & 11.20 &
 72.44$\pm$0.74 & 72.44$\pm$0.74 \\
Pal4-28 & S426 &  35.9 & 19.87 & 0.91 & 1500 & 11222.1192 &  5.89 &
 72.20$\pm$1.31 & 72.20$\pm$1.31 \\
Pal4-30 & S276 &  99.7 & 19.89 & 0.90 & 1800 & 11248.0166 &  9.50 &
 71.33$\pm$0.86 & 71.33$\pm$0.86 \\
Pal4-31 & S315 &   7.5 & 19.89 & 0.93 & 1500 & 11222.1982 & 10.08 &
 72.38$\pm$0.81 & 72.38$\pm$0.81 \\
\hline \hline
\end{tabular}
\label{tab:radvel}
\end{table*}

\section{The systemic velocity and the velocity dispersion}
\label{sec:specresults}
\begin{figure}
\includegraphics[width=84mm]{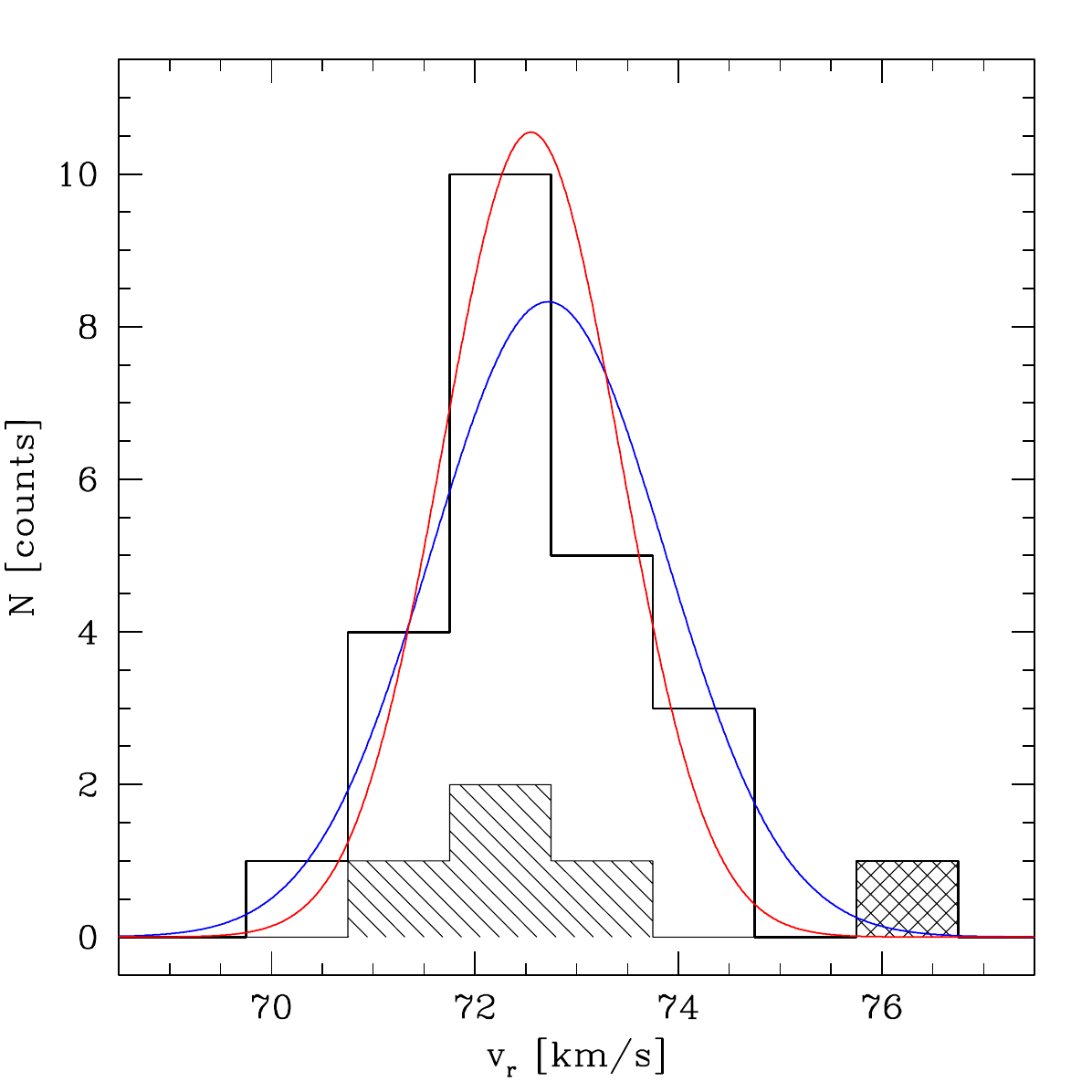}
\caption{Histogram of radial velocities for all 24 sample stars. The hashed
areas correspond to AGB stars, the cross-hashed area corresponds to star\,12
at
$\sim76\kms$.
The blue and red curves are the maximum-likelihood Gaussian representations of
\emph{intrinsic} velocity distribution for the total sample of 24 stars and
for
the sample without star\,12, respectively. The sigmas
of the Gaussian are the velocity dispersions as derived using the
\citet{1993ASPC...50..357P} method.}
\label{fig:velhist}
\end{figure}

Table\,\ref{tab:radvel} summarizes the results of our radial velocity
measurements for Pal\,4 member stars. Columns (1)--(10) of this table record
the names of
each program star \citep[second column from identification
by][]{2005PASP..117...37S},
distance from the cluster center, $V$ magnitude, $(B-V)$ color \citep[both
from][]{2005PASP..117...37S}, HIRES exposure time, the heliocentric Julian
date
of the observation, the Tonry \& Davis $R_{\rm TD}$ value, the heliocentric
radial velocity, and the
error-weighted mean velocity. Six of the stars in our Pal\,4 sample were
observed twice, and two stars were observed three times. For most stars the
difference
in radial velocity between the individual measurements is below 1$\kms$.
Two stars show a larger discrepancy of 1.65$\kms$ (Pal4-10) and 4.51$\kms$
(Pal4-12, a likely AGB star), potentially due to binarity. For the latter, the
two velocity measurements differ by
more than $5\sigma$ and the mean of the two measurements stands out in the 
velocity distribution (see Fig.\,\ref{fig:velhist}). This suggests that the
star
should probably be excluded as an outlier. Nevertheless, as its mean velocity
is still marginally
consistent with the velocity distribution (see below), we will present our
kinematical analysis with and
without this star (named in the following `star\,12').

The mean heliocentric radial velocity and velocity dispersion of Pal\,4
were calculated using the maximum likelihood method of
\citet{1993ASPC...50..357P}. For
details about the method see also sec.\,3.2 of \citet{2009MNRAS.396.2051B}.
Using the 23 clean member stars from Table\,\ref{tab:radvel} (i.e. excluding
star 12), we obtain a mean
cluster velocity of  $v_r=72.55\pm0.22\kms$ and an intrinsic velocity
dispersion of $\sigma=0.87\pm0.18\kms$. When including star\,12, the mean
velocity is $v_r=72.72\pm0.27\kms$ and the velocity dispersion rises to
$\sigma=1.15\pm0.20\kms$. The cluster's mean radial velocity is consistent
with the determination by \citet{1992AJ....104..164A}, $v_r=74\pm1\kms$.

\begin{figure}
\includegraphics[width=84mm]{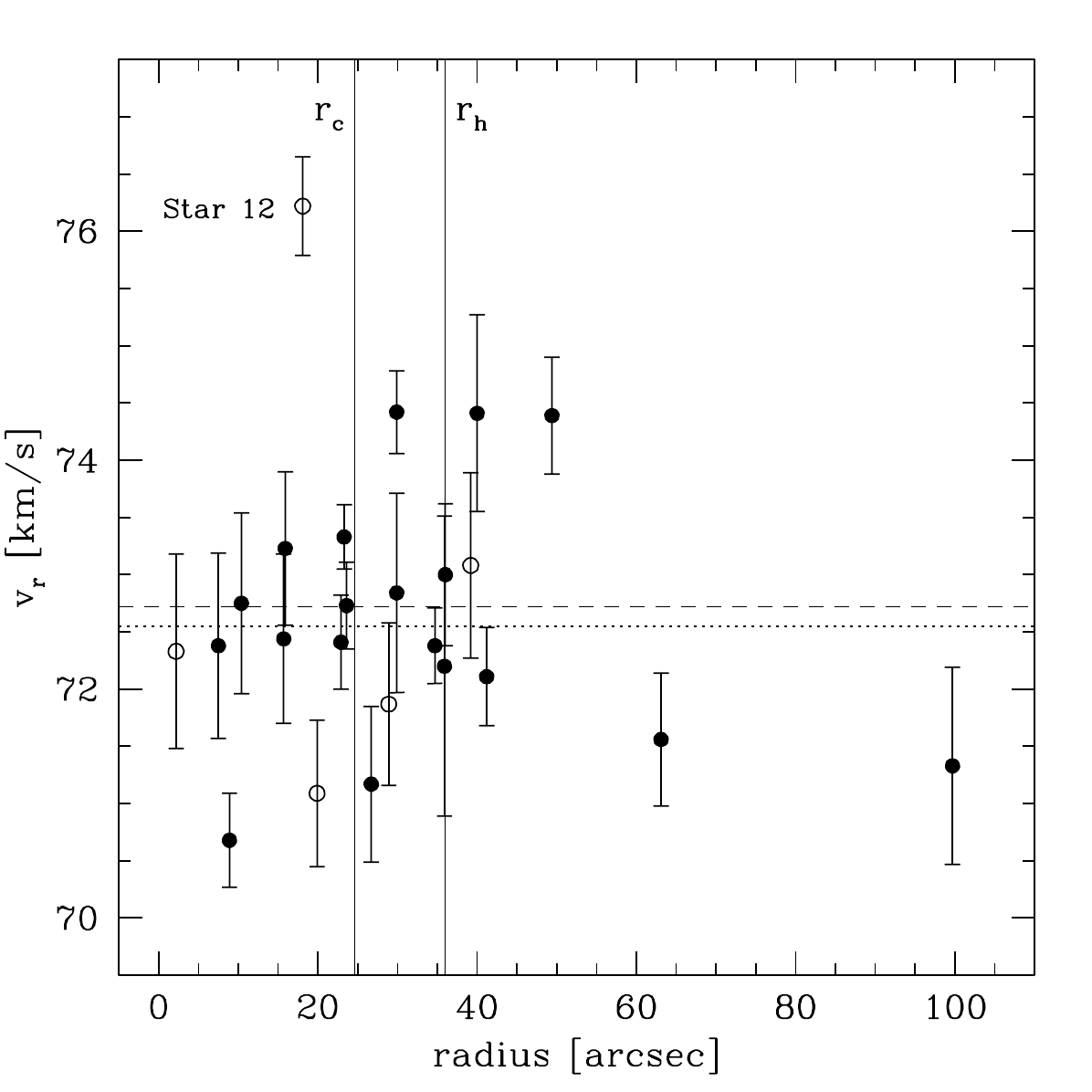}
\caption{Radial distribution of stars with velocity measurements in
Table\,\ref{tab:radvel}. The open symbols mark probable AGB stars. The
horizontal dotted line marks Pal\,4's error-weighted mean systemic velocity
without
star\,12, and the dashed line the velocity including star\,12. The core
and half-light radii are indicated by the vertical lines.}
\label{fig:velrad}
\end{figure}

\begin{figure}
\includegraphics[width=84mm]{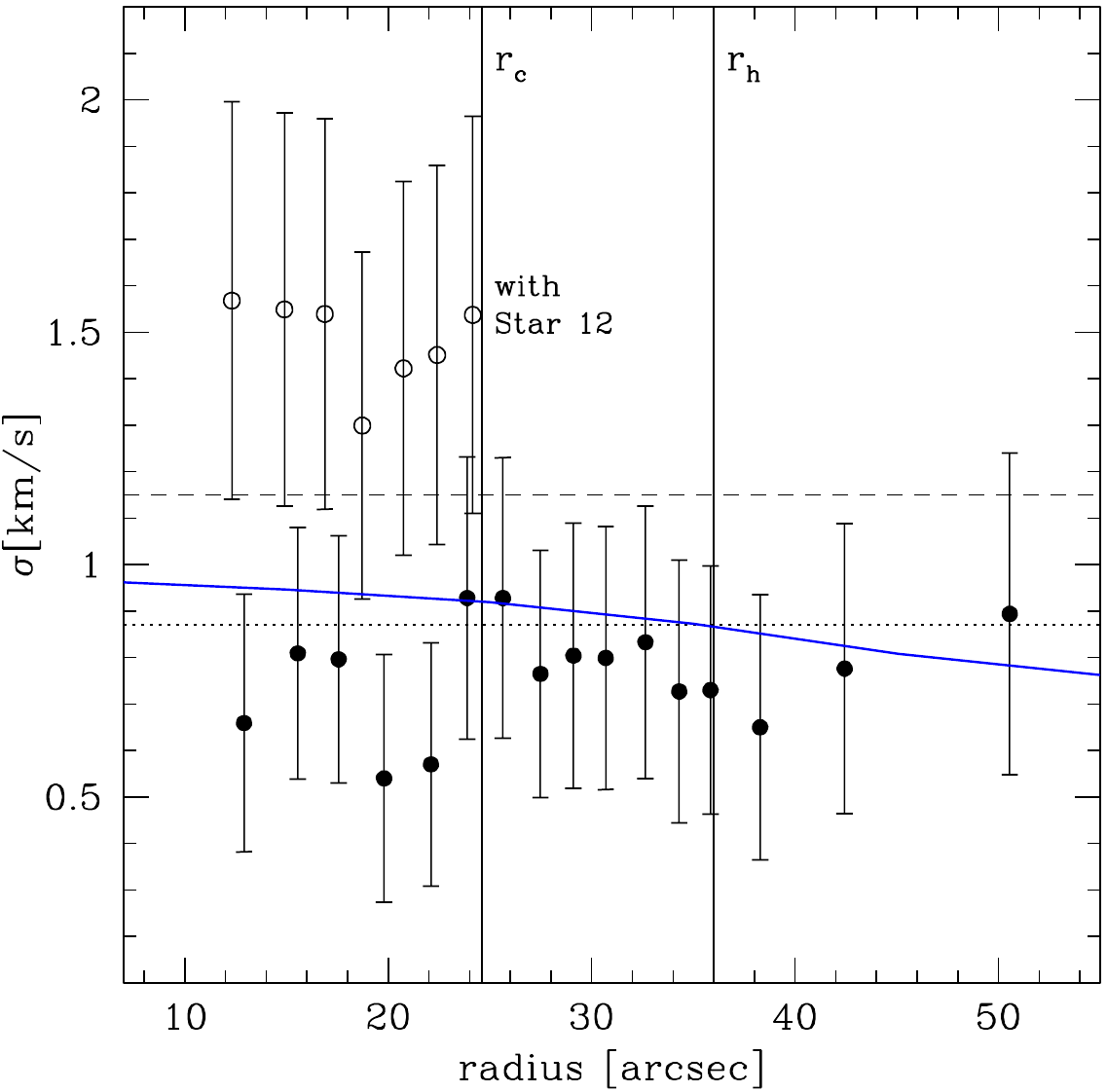}
\caption{Velocity dispersion profile of Pal\,4 using running bins with eight
stars in each bin. The black filled symbols denote the velocity dispersion
without
star\,12. The open symbols denote these bins where star\,12 was included.
The dashed and dotted horizontal lines are the average dispersion values
if star\,12 is included or excluded, respectively. The vertical lines are
the core and half-light radii. Shown as blue solid curve is the dispersion 
profile expected in Newtonian dynamics for a cluster mass of
$2.98\times10^{4}\msun$ 
and assuming that mass follows the light of the best-fitting
\citet{1966AJ.....71...64K} 
model derived in Section~\ref{sec:surfacebrightness}.}
\label{fig:disprad}
\end{figure}

Fig.\,\ref{fig:velhist} shows the distribution of radial velocities of the
24 cluster members (open histogram). The curves show the maximum-likelihood
Gaussian representations of the intrinsic velocity distribution (with and
without star\,12) using the above values for
$v_r$ and $\sigma$. As can be seen, the observed radial velocity distribution
is well approximated by a Gaussian except for the outlier star\,12. For a
Gaussian
distribution and a sample of 24 stars, one would expect to find a star that
is, like star 12, about 3$\sigma$ away from the mean in only 5 per cent  of all
cases.

In Fig.\,\ref{fig:velrad} we show the radial distribution of our measured
velocities
(star\,12 is labeled). The cluster's mean velocity is marked by the dotted
(without star\,12) and dashed (with star\,12) horizontal line. One third of
the
24
sample stars are located at radii equal to or greater than the half-light
radius.
Thus,
the measured velocity dispersion is only slightly biased towards the central
value. In this plot no clear trend of a decreasing or increasing velocity
dispersion with radius is seen. However, our sampling beyond 50 arcsec radius
is very sparse with only two measured velocities. Nevertheless, we derived
the line-of-sight velocity dispersion profile with running radial bins, each
bin containing eight stars. Fig.\,\ref{fig:disprad} shows the resulting
velocity
dispersion profile. Within a radius of up to 24 arcsec we derived the velocity
dispersion either excluding star\,12 or including star\,12. For the case
excluding star\,12,
we can see a flat velocity dispersion profile that is in good agreement with
the expectation 
from a single-mass, non-mass-segregated King model that is overplotted. When
including star\,12 one might argue for a declining velocity dispersion
profile.

\begin{figure}
\includegraphics[width=84mm]{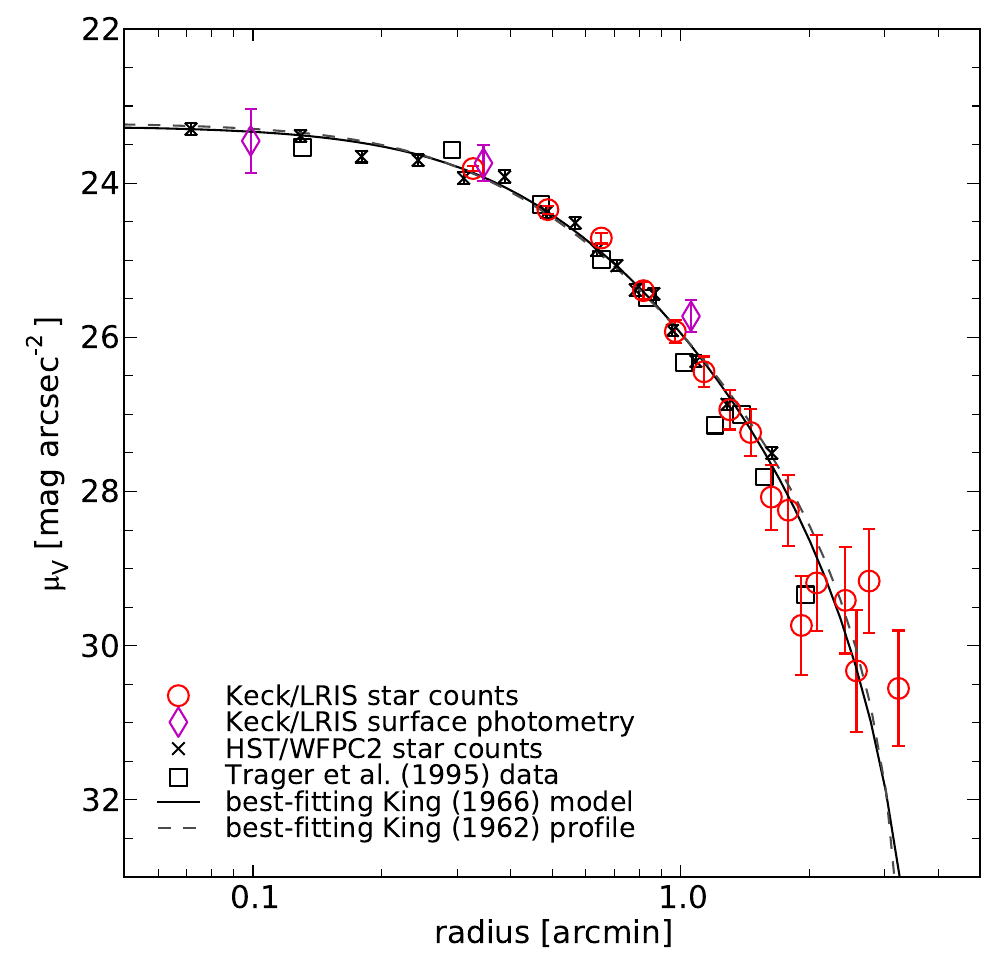}
\includegraphics[width=84mm]{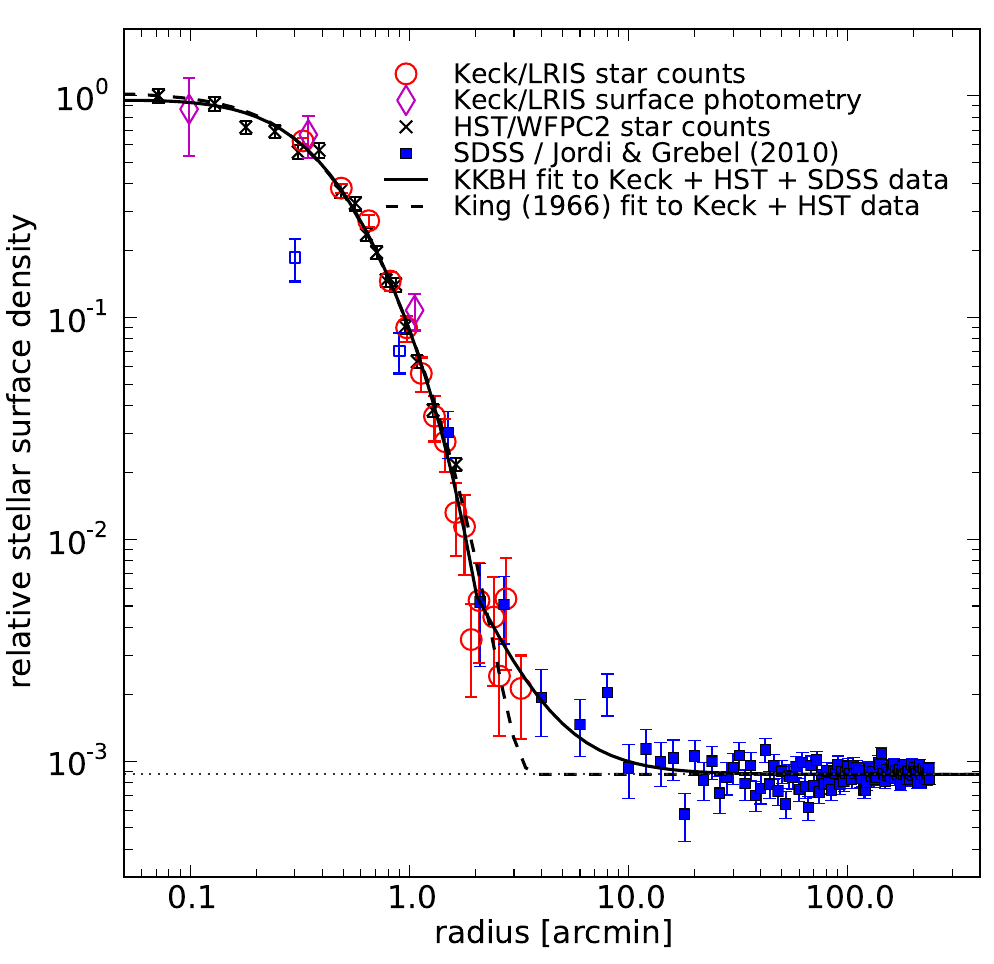}
\caption{Top: The surface brightness profile of Pal\,4. Our LRIS data are
represented by filled red circles (derived from star counts) and filled
magenta diamonds (from direct surface photometry), the WFPC2 star counts are
represented by black crosses. Open squares show the
\citet{1995AJ....109..218T} data based on star counts on photographic plates.
The best-fitting \citet{1966AJ.....71...64K} model to the Keck and \textit{HST} data is
shown as solid curve, the best-fitting \citet{1962AJ.....67..471K} profile is
shown as dashed curve. Bottom: The cluster's surface density profile
(normalized to the innermost point). In order to display the profile on a
logarithmic scale we added a virtual background level, indicated as dotted
horizontal line. As in the upper panel, red circles and magenta diamonds
represent the Keck data and black crosses represent the \textit{HST} WFPC2 data. Blue
squares represent the SDSS-based profile derived by
\citet{2010A&A...522A..71J}. The two innermost data points of the SDSS-based
profile (shown as open squares) were excluded, because they are systematically
low due to crowding. The best-fitting KKBH profile to the combined dataset is
shown as solid curve. For comparison, the dashed curve represents the
best-fitting \citet{1966AJ.....71...64K} model from the top panel. Pal\,4
shows a clearly enhanced stellar density at radii $>3$\,arcmin.}
\label{fig:surfacebrightnessprofile}
\end{figure}

\begin{table*}
\caption[]{Structural parameters of Pal\,4. A distance of $102.8 \pm 2.4$~kpc
(Section~\ref{sec:photresults:age}) was adopted and all literature values
dependent on distance were recalculated using this distance. In calculating the total luminosity L$_V$, we used a $V$-band extinction of A$_V$=$3.1\times E(B-V)=0.07$\,mag (Section~\ref{sec:photresults:age}) and $M_{V,\odot}=4.83$\,mag \citep{1998gaas.book.....B}.}
\begin{tabular}{|l|r|l|r|l|}
\hline \hline
\multicolumn{3}{c}{best-fitting King (1966) model} & \multicolumn{2}{c}{King
(1966) model of} \\ 
 & &  & \multicolumn{2}{c}{\citet{2005ApJS..161..304M}} \\
\hline
central surface brightness $\mu_{V,0}$ & $23.26\pm0.06$ & mag\,arcsec$^{-2}$ &
$23.01^{+0.26}_{-0.22}$& mag\,arcsec$^{-2}$ \\
core radius $r_c$ & $0.43\pm0.03$ & arcmin & $0.33^{+0.05}_{-0.04}$ & arcmin
\\ 
& $13.0\pm0.8$& pc & $9.8^{+1.4}_{-1.3}$& pc \\ 
tidal radius $r_t$ & $3.90\pm0.20$& arcmin & $3.30\pm0.23$& arcmin\\ 
& $116.7\pm6.6$& pc  & $98.6\pm7.2$& pc \\ 
concentration $c$ &  $0.96\pm0.04$ & & $0.93\pm0.1$ &\\ 
2d half-light radius $r_{h}$ & $0.62\pm0.03$& arcmin & $0.51^{+0.03}_{-0.02}$&
arcmin \\ 
& $18.4\pm1.1$& pc  & $15.3^{+0.9}_{-0.8}$& pc  \\
apparent magnitude $V$ & $14.23\pm0.03$ & mag  & $14.33^{+0.06}_{-0.03}$& mag  \\

total luminosity L$_V$ & $19600\pm1100$ & L$_\odot$ & $17900^{+1000}_{-1300}$& L$_\odot$
\\ 
& & & & \\
\hline
\multicolumn{3}{c}{best-fitting \citet{1962AJ.....67..471K} profile} &
\multicolumn{2}{c}{\citet{1962AJ.....67..471K} profile of} \\
 & & & \multicolumn{2}{c}{Jordi \& Grebel (2010)} \\
\hline
central surface brightness $\mu_{V,0}$ & $22.96\pm0.05$& mag\,arcsec$^{-2}$ &
-- &\\
core radius $r_c$ & $0.39\pm0.02$ & arcmin & $0.26\pm0.10$ & arcmin\\
 & $11.7\pm0.6$ & pc  & $7.8\pm3.0$& pc \\
tidal radius $r_t$ & $3.46\pm0.16$& arcmin & $5.30\pm0.65$ & arcmin\\
&  $103.6\pm5.4$& pc  & $158\pm20$& pc  \\
2d half-light radius $r_{h}$ &  $0.63\pm0.03$&arcmin& $0.62\pm0.24$&arcmin\\
& $18.8\pm1.0$&pc& $18.7\pm7.2$&pc \\\medskip
& & & & \\
\hline 
\multicolumn{3}{c}{best-fitting KKBH profile} & & \\
\multicolumn{3}{c}{to combined LRIS, WFPC2 and Jordi \& Grebel (2010) data} &
& \\
\hline 
central surface brightness $\mu_{V,0}$ & $22.88\pm0.17$ & mag\,arcsec$^{-2}$ &
&\\
inner power-law slope $\gamma$ & $-0.04\pm0.13$ & & &\\
core radius $R_c$ & $0.44\pm0.04$& arcmin & &\\
 & $13.1\pm0.3$& pc  & &\\
edge radius $R_t$& $2.77\pm0.12$& arcmin &  &\\
&  $82.9\pm1.9$& pc  & & \\
turn-over parameter $\mu$ & $0.72\pm0.05$ & & & \\
outer power-law slope $\eta$ & $2.3\pm0.6$ & & &\\
\hline \hline
\end{tabular}
\label{tab:surfacebrightness}
\end{table*}

\section{Photometric results}
\label{sec:photresults}

\subsection{Surface brightness profile \& structural parameters}
\label{sec:surfacebrightness}
In the literature there are only few surface brightness profiles and
derivations of the structural parameters of Pal\,4. As mentioned in the
introduction, in a search for extra-tidal features \citet{2003AJ....126..803S}
used deep wide-field imaging to study the stellar density distribution around
Pal\,4. Unfortunately, they did not derive a density profile or the cluster's
structural parameters, but adopted the structural parameters from the
\citet{1996AJ....112.1487H} catalog. This catalog in its 2003 version quoted
the structural parameters derived by \citet{1995AJ....109..218T} from a
compilation of surface photometry. In its updated 2010 version, the
\citet{1996AJ....112.1487H} catalog refers to the reanalysis of the
\citet{1995AJ....109..218T} data presented by \citet{2005ApJS..161..304M}.
Recently, in a search for tidal tails around Galactic GCs,
\citet{2010A&A...522A..71J} derived surface density profiles for 17 GCs,
including Pal\,4. These are based on star counts in the Sloan Digital Sky
Survey \citep[SDSS DR7;][]{2009ApJS..182..543A} catalog and the PSF-fitting
photometry of SDSS imaging of the inner regions of Galactic GCs by
\citet{2008ApJS..179..326A}. However, the authors note that Pal\,4 is the most
distant GC in their sample and thus the sample includes only stars on the
upper RGB. Moreover the cluster's large distance and the relatively bright
limiting magnitude and low spatial resolution of the SDSS make crowding an
issue, at least in the cluster's inner region (r$\la$1\,arcmin).  

We therefore used our Keck LRIS photometry to measure the structural
parameters for Pal\,4. The point-source catalog from our LRIS images covers an
area of 42.8\,arcmin$^{2}$ and contains 777 objects, after excluding stars
fainter than $V$ = 24.5\,mag to minimize photometric incompleteness. Star
counts based on these data were then combined with surface photometry for the
innermost regions to construct a composite $V$-band surface brightness profile
for the cluster, using the approach described in \citet{1992AJ....103..857F}.
The upper panel of Fig.\,\ref{fig:surfacebrightnessprofile} shows the
resulting surface brightness profiles (filled red circles) and additionally
three data points from direct surface photometry on the $V$-band image (filled
magenta diamonds). As the deeper WFPC2 data sample a much greater number of
stars in the cluster's center, we also included a surface brightness profile
derived from star counts in the WFPC2 catalog and the $V$-band magnitudes that
\textsc{HSTPHOT} calculates based on the \citet{1995PASP..107.1065H} WFPC2 to $UBVRI$
transformations. We included stars down to 27\,mag in F555W and corrected the
star counts and flux for the radially varying completeness. The resulting
profile is shown as black crosses in
Fig.\,\ref{fig:surfacebrightnessprofile}; due to the inhomogeneous geometric
coverage of the WFPC2 catalog, we define radial bins by the requirement that
they hold equal numbers of stars. Thus, the Poissonian errorbars on the data
points remain constant, while their radial spacing varies. 

Both surface
brightness profiles agree very well and also show good agreement with the
\citet{1995AJ....109..218T} surface brightness data, which are shown for
comparison as open black squares.
The figure also shows the best-fitting \citet[][solid
curve]{1966AJ.....71...64K} model to our LRIS and WFPC2 data, which yields
a central surface brightness of $\mu_{V,0}=23.26\pm0.06$\,mag\,arcsec$^{-2}$,
a core radius of $r_c=0.43\pm0.03$\,arcmin and a tidal radius of
$r_t=3.90\pm0.20$\,arcmin, corresponding to a concentration of
$c=\mathrm{log}(r_t/r_c)=0.96\pm0.04$ and a (two-dimensional) half-light
radius of $r_{h}=0.62\pm0.03$\,arcmin. For comparability, we also fitted a
\citet{1962AJ.....67..471K} profile to our data (dashed curve), which yields
core and tidal radii of $r_c=0.39\pm0.02$ and $r_t=3.46\pm0.16$\,arcmin and a
central surface brightness of $22.96\pm0.05$\,mag\,arcsec$^{-2}$ and
reproduces the observations marginally worse in terms of the minimum $\chi^2$.
Table~\ref{tab:surfacebrightness} summarizes our fit results and shows also
literature values for comparison. Our best-fitting \citet{1966AJ.....71...64K}
model is somewhat more extended than the one derived by
\citet{2005ApJS..161..304M}, but otherwise is in good agreement with the
latter in terms of central surface brightness, concentration and integrated
total luminosity. Comparing our best-fitting \citet{1962AJ.....67..471K}
profile to that of \citet{2010A&A...522A..71J}, we find that the latter is
more extended and diffuse. This is consistent with the SDSS data
underestimating stellar density in the cluster's center due to crowding as we
will see below. 

\citet{2003AJ....126..803S} noted an excess of stars beyond the cluster's
formal tidal radius, for which they adopted $r_t=3.33$\,arcmin. As our Keck
data reach out to a radius of only $\sim\,3.2$\,arcmin, we combine our profile
with the SDSS-based profile of \citet{2010A&A...522A..71J}. We scaled their
background-corrected surface density profile (K. Jordi, private communication)
to match the Keck data in the radial range of $1.5-3.2$\,arcmin, by
interpolating the Keck data to the radii of the SDSS data points and requiring
that the median ratio of the two profiles in the overlapping region be one.
The merged profile is shown in the lower panel of
Fig.\,\ref{fig:surfacebrightnessprofile}. As before, diamonds and circles
represent the Keck profile, crosses represent the WFPC2 profile. Blue squares
represent the SDSS profile. As the SDSS data reach beyond the tidal radius the
background-corrected stellar density in individual radial bins can scatter
below zero. For the purpose of plotting the profile on a logarithmic scale, we
therefore added an artificial background level (shown as dotted horizontal
line). The two innermost points of the SDSS data, shown as open squares,
deviate from the Keck and WFPC2 profile reflecting the crowding in the SDSS
data and we excluded them in our analysis. The dashed line represents the
best-fitting \citet{1966AJ.....71...64K} model from above, and it is obvious
that the observed density at large radii falls off less steeply than this
model or any other similarly truncated model. We fitted the combined profile
with a \citet[][KKBH]{2010MNRAS.407.2241K} template. These templates were
designed to fit surface density profiles of GCs out to large cluster radii
based on fits to a suite of N-body simulations of Galactic GCs on various
orbits. They are a modification of the \citet{1962AJ.....67..471K} profile
including a term for a non-flat core and a term for tidal debris. The
best-fitting KKBH profile, shown as solid line in the lower panel of
Fig.\,\ref{fig:surfacebrightnessprofile} is found for core and edge radii of
$R_c=0.44\pm0.04$\,arcmin and $R_t=2.77\pm0.12$\,arcmin, a core power-law
slope of $\gamma=0\pm0.1$ and an outer power-law slope of $\eta=2.3\pm0.6$
that becomes dominant at $\mu R_t=2.00\pm0.15$\,arcmin. The shallow slope at
large cluster radii may indicate that the cluster is in an orbital phase close
to its apogalacticon, although projection effects may play a role in the appearance 
of the outer part of the density profile. \citet{2010MNRAS.407.2241K} find that the surface
density profiles of star clusters, as seen in projection onto their orbital planes, are influenced by the tidal debris in this orbital phase: while the slope at large cluster radii, $\eta$, is about 4-5 in
most orbital phases, it can reach values of 1-2 in apogalacticon due to
orbital compression of the cluster and its tidal tails. 

For our following analysis, we will adopt the best-fitting
\citet{1966AJ.....71...64K} model as the cluster's density profile and come
back to the influence of tidal debris in Section~\ref{sec:disc:effectsof}.

\subsection{Age determination}
\label{sec:photresults:age}
To derive the cluster's age, we determined the isochrone that best reproduces
the locus of the principal evolutionary sequences from a subset of isochrones
of
the Dartmouth Stellar Evolution Database \citep{2008ApJS..178...89D}. Based on
the chemical composition derived by \citet{2010A&A...517A..59K} from coadded
high-resolution spectra of red giants, we adopted [Fe/H]=$-1.41$~dex and an
$\alpha$-enhancement of $+0.4$~dex. We determined the best-fitting isochrone
using a robust direct fit \citep[similar to][]{1999AJ....117..247S}, to the
color-magnitude data. As the subgiant branch is almost horizontal in the CMD,
even in the F814W vs. F555W-F814W plane (used by \citet{1999AJ....117..247S}
for
that reason), a minimization in one dimension (interpreting the isochrone as
`color as a function of magnitude' and comparing the separation in color of
each
star to the color uncertainty in that magnitude range) runs into problems.
Therefore, we employed a $\chi^2$ minimization in the (F555W, F814W)-plane,
where
the uncertainties in both dimensions are uncorrelated, and minimized the
squared sum 
of 2d-distances of each star to the isochrone. To be less sensitive to
outliers, 
instead of $\chi^2$, a robust metric that saturates at 5 $\sigma$ was used. 
Distance and age were varied as free parameters, with the latter ranging from
8~Gyr 
to 15~Gyr in steps of 0.5~Gyr. We adopted a reddening of
$\mathrm{E}(B$-$V)=0.023{\rm\,mag}$ 
estimated from Galactic dust emission maps\footnote{Obtained from
http://irsa.ipac.caltech.edu/applications/DUST/} and filter-specific
extinction
to reddening ratios of A$_\mathrm{555W}/\mathrm{E}(B$-$V)$=3.252 and
A$_\mathrm{F814W}/\mathrm{E}(B$-$V)$=1.948, taken from table~6 of
\citet{1998ApJ...500..525S}. From
this, we obtained a best-fitting age of $11 \pm 1$~Gyr and an 
extinction-corrected distance modulus of $20.06 \pm 0.05{\rm\,mag}$. This
places
the cluster at a distance of $102.8\pm2.4$~kpc from the Sun. This is slightly
closer
than the 109.2~kpc derived by \citet[edition 2010]{1996AJ....112.1487H} from
the mean observed $V$-band magnitude of horizontal branch stars from
\citet{1999AJ....117..247S}, but well within the range of other previous
distance determinations of 100~kpc \citep{1958ApJ...127..527B}, $105\pm5$~kpc
\citep{1986ApJ...303..216C} and 104~kpc \citep{2000ApJS..129..315V}. The age
estimate is
consistent with Pal\,4 being part of the young halo population and
$\sim\,1.5$--2~Gyrs younger than `classical', old GCs, as also suggested by
the
differential analysis relative to M\,5 by \citet{1999AJ....117..247S} and
\citet{2000ApJS..129..315V}.

\subsection{Mass function}
\label{sec:photresults:massfunction}

\begin{figure}
\includegraphics[width=84mm]{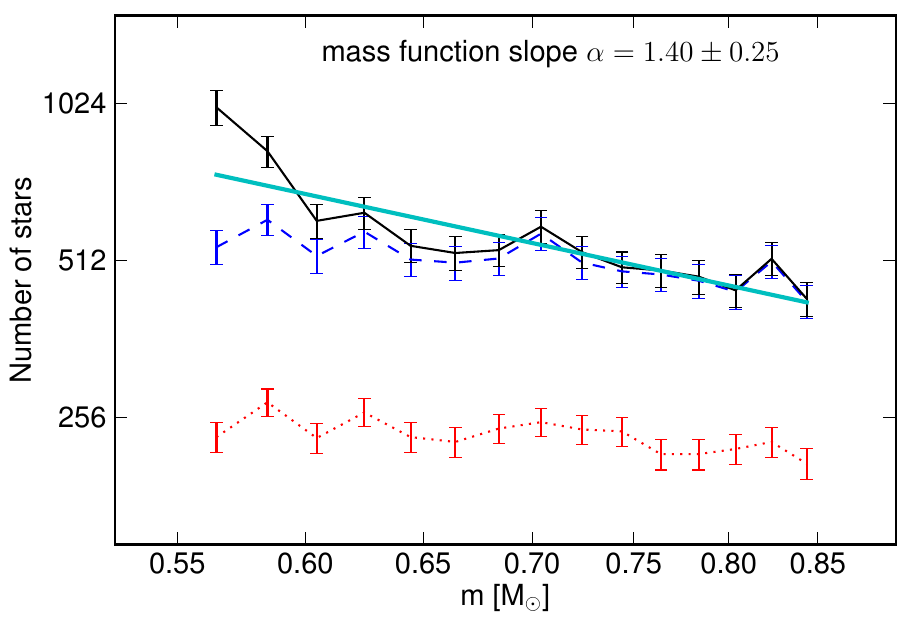}
\caption{Mass function and power-law fit. The red dotted curve shows the
number
of observed stars per mass interval, errorbars represent the Poissonian errors
on the star counts. The blue dashed curve represents the counts corrected for
the missing area coverage, the black solid curve represents the counts
additionally
corrected for photometric completeness. The cyan line gives the best-fitting
power
law.}
\label{fig:massfunc}
\end{figure}
\begin{figure}
\includegraphics[width=82mm]{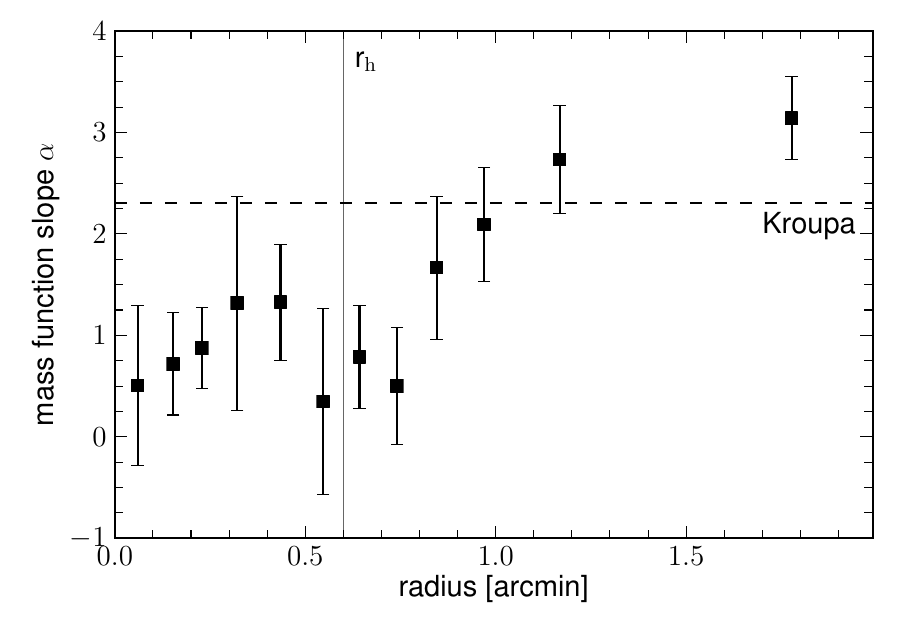}
\includegraphics[width=82mm]{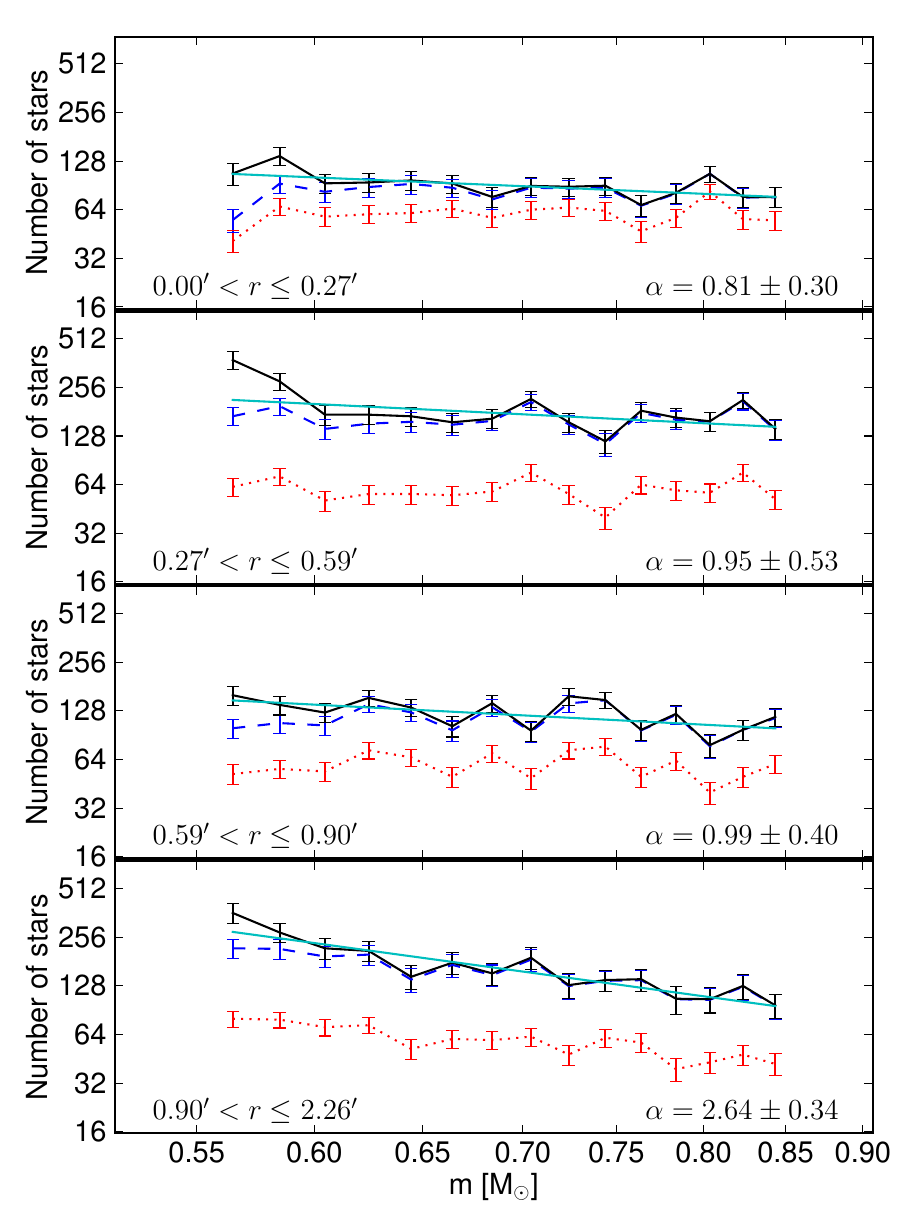}
\caption{Top: The best-fitting mass function
slope $\alpha$ in radial bins containing each one twelfth of the observed
stars. Bottom: The mass function in radial bins containing one fourth of the
observed stars each. Dotted red curves represent the number of observed  stars
per mass interval, errorbars represent the Poissonian errors on the star
counts. Blue dashed
curves are the counts corrected for the missing area coverage in the given
radial range. Black solid curves are additionally corrected for photometric
completeness.
Cyan lines represent the best-fitting power law functions to the
completeness-corrected counts. The radial ranges and best-fitting
power-law slopes are reported at the bottom of each panel. }
\label{fig:masssegregation}
\end{figure}

We determined the stellar mass function in the cluster in the mass range
$0.55\le \mathrm{M}/\msun\le 0.85$, corresponding to stars from the
tip of the RGB down to the 50 per cent completeness limit in the cluster's core at
the
faint end ($17.9{\rm\,mag}\la{\rm F555W}\la27.6{\rm\,mag}$). We rejected stars
that deviated in color from the locus of the isochrone by more than
3$\sigma_{col}$, where $\sigma_{col}$ is the color uncertainty derived from
the
artificial star results in the corresponding region of the CMD. To avoid
rejecting RGB stars, whose scatter around the isochrone is slightly larger
than
expected purely from photometric uncertainties, we additionally allowed for an
intrinsic color spread of $0.02{\rm\,mag}$. This selection removed likely
foreground stars, blue stragglers and horizontal branch stars (see
Fig.~\ref{fig:CMD}). We then assigned to each of the remaining stars a mass
based on the isochrone, by interpolating the masses tabulated in the isochrone
to the star's measured F555W magnitude.

At the faint end, crowding affects the photometry and thus the completeness
varies slightly with stellar density, or distance from the cluster center.
Moreover the geometric coverage of the WFPC2
photometry as a function of radius is very inhomogeneous (see
Fig.~\ref{fig:completeness}). 
Therefore, we subdivided our photometric catalog into n radial bins around the
cluster center, chosen
such that each bin contains one n-th of the observed stars. This is optimal in
terms of the Poissonian errors on the star counts, both of the observed stars
and of the artificial stars, as the latter were distributed on the sky
similarly
to the observed stars. The number of radial subdivisions has to be chosen
large
enough such that completeness and stellar density are approximately constant
within each annulus, because otherwise correcting for completeness would
bias the results. In practice, we increased the number of bins, n, until the
derived mass function slope and cluster mass
(Section~\ref{sec:photresults:totalmass}) did not vary any more with n. This
was
the case for n\,$\ge$\,33 and we chose n\,$=$\,36 radial bins for the final
analysis. In each of these annuli, stars were counted in 12 linearly spaced
mass
bins (of width $\sim0.025~\msun$). The counts were corrected for
the missing area coverage and for photometric completeness in that radial
range.
Counts from the individual annuli were then summed and fit with a power law of
the form $dN/dm\propto m^{-\alpha}$. From this, we obtained a mass function
slope of $\alpha = 1.4 \pm 0.25$ (Fig.~\ref{fig:massfunc}). This present-day
mass function is significantly shallower than a \citet{2001MNRAS.322..231K}
IMF (with $\alpha=2.3$ in this range of masses) and is similar to the
mass function in other Galactic GCs
\citep[e.g][]{2007ApJ...656L..65D,2009AJ....137.4586J,2010AJ....139..476P}.

\subsection{Mass segregation}
\label{sec:photresults:masssegregation}
To test for mass segregation, we derived the mass function as a function of
radius. As the individual 36 radial annuli contain only $\sim$120 stars each,
deriving the mass function in each of them would produce very noisy results.
It is thus necessary to bin several of these annuli -- after the
completeness-corrected counts have been obtained in each annulus individually.
As a compromise between signal to noise and radial resolution, we show two
different binning schemes: The top panel of Fig.~\ref{fig:masssegregation}
shows the best-fitting mass function slopes derived in radial bins containing
each one twelfth of the observed stars. The bottom panel of the same figure
shows the mass functions and power-law fits obtained in bins containing each
one fourth of the observed stars. It is obvious that the mass function
steepens with increasing radius, from $\alpha\la1$ inside $r\la1.3\times r_h$
to $\alpha\ga$2.3 at the largest observed radii.

\subsection{Total mass}
\label{sec:photresults:totalmass}
In the mass range between $0.55\le \mathrm{M}/\msun\le 0.85$, we
measure a stellar mass of $5960\pm110\,\msun$ within the radius 
covered by the WFPC2 pointing, $r<2.26$\,arcmin. We do not correct for the
mass contained in blue stragglers and horizontal branch stars that fall
outside of our color selection. It is negligible due to their low number
($\sim~20$ of each species in our pointing) and we estimate their contribution
to be $\la 0.2$ per cent of the total cluster mass.

Assuming the measured mass function slope of $\alpha=1.40 \pm 0.25$ to hold
down to
0.5~$\msun$ and adopting a \citet{2001MNRAS.322..231K}
mass function, with $\alpha=1.3$, for masses $0.08\le
M/\msun\le 0.5$, and $\alpha=0.3$ for masses $0.01\le
M/\msun\le 0.08$, the extrapolated stellar mass in the
mass range $0.01\le \mathrm{M}/\msun\le 0.85$ is $14500\pm1300\,\msun$.

To account for the mass contributed by the remnants of higher-mass stars, we
assume our observed slope $\alpha$ to hold up to 1.0~$\msun$ and
above that a high-mass Kroupa slope of $\alpha=2.3$, and extrapolate the mass
function to
60\,$\msun$. We follow the prescription of
\citet{2011AJ....142...36G}, assuming stars with initial masses $0.85\le \mathrm{M} \le
8\msun$ to have formed 0.6\,$\msun$ white dwarfs, and
stars with initial masses $8\le \mathrm{M} \le 60\msun$ to have formed
neutron stars of 1\,$\msun$. The extrapolation yields a mass in white dwarfs
of $\mathrm{M}_\mathrm{WD}$=$8900\pm800\,\msun$ and a mass in neutron stars of
$\mathrm{M}_\mathrm{NS}=800\pm70\,\msun$. In clusters with masses of several times
$10^4\,\msun$, neutron stars are expected to escape the cluster due to their
high initial kick velocities, while virtually all white dwarfs are expected to
be retained in the cluster \citep{2009A&A...507.1409K}. We therefore adopt
$\mathrm{M}_\mathrm{WD}$=$8900\pm800\,\msun$ as the mass of stellar remnants.

Based on the best-fitting \citet{1966AJ.....71...64K} density profile, and
approximating that mass
follows light, $98.3\pm0.4$ per cent of the cluster's mass lies within
$r=2.26$\,arcmin. 
Extrapolating out to the tidal radius, the total mass of Pal\,4 amounts to
$\mathrm{M}_\mathrm{phot}=29800\pm800\,\msun$ including the corrections for low-mass
stars and stellar remnants. We note that the uncertainty of the total mass is
smaller than the individual uncertainties of the extrapolated high- and
low-mass contributions because correlations were fully propagated. These
correlations arise from the requirement that the mass function be continuous.
As a steeper (shallower) mass function will have more (less) mass in low-mass
stars and less (more) mass in high-mass stars and stellar remnants, the
uncertainties of the two terms are anti-correlated.

With this mass and the total luminosity derived from the best-fitting
\citet{1966AJ.....71...64K} model, the photometric mass to light ratio of the
cluster is M$_\mathrm{phot}$/L$_{V}=1.52\pm0.09\,\msun\lsun^{-1}$.

To obtain a conservative lower limit on the photometric mass of the cluster,
we
follow \citet{2009AJ....137.4586J}, assuming the cluster to be significantly
depleted
in low-mass stars with a declining mass function with $\alpha=-1.0$ for masses
$0.01\le \mathrm{M}/\msun\le 0.5$. For this hypothetical case, the
extrapolation towards lower masses, inclusion of white dwarfs and
extrapolation out to the tidal radius yields a
total cluster mass of M$_\mathrm{decl,phot}=20100\pm600\,\msun$. 

\section{Discussion}
\begin{figure}
\includegraphics[width=84mm]{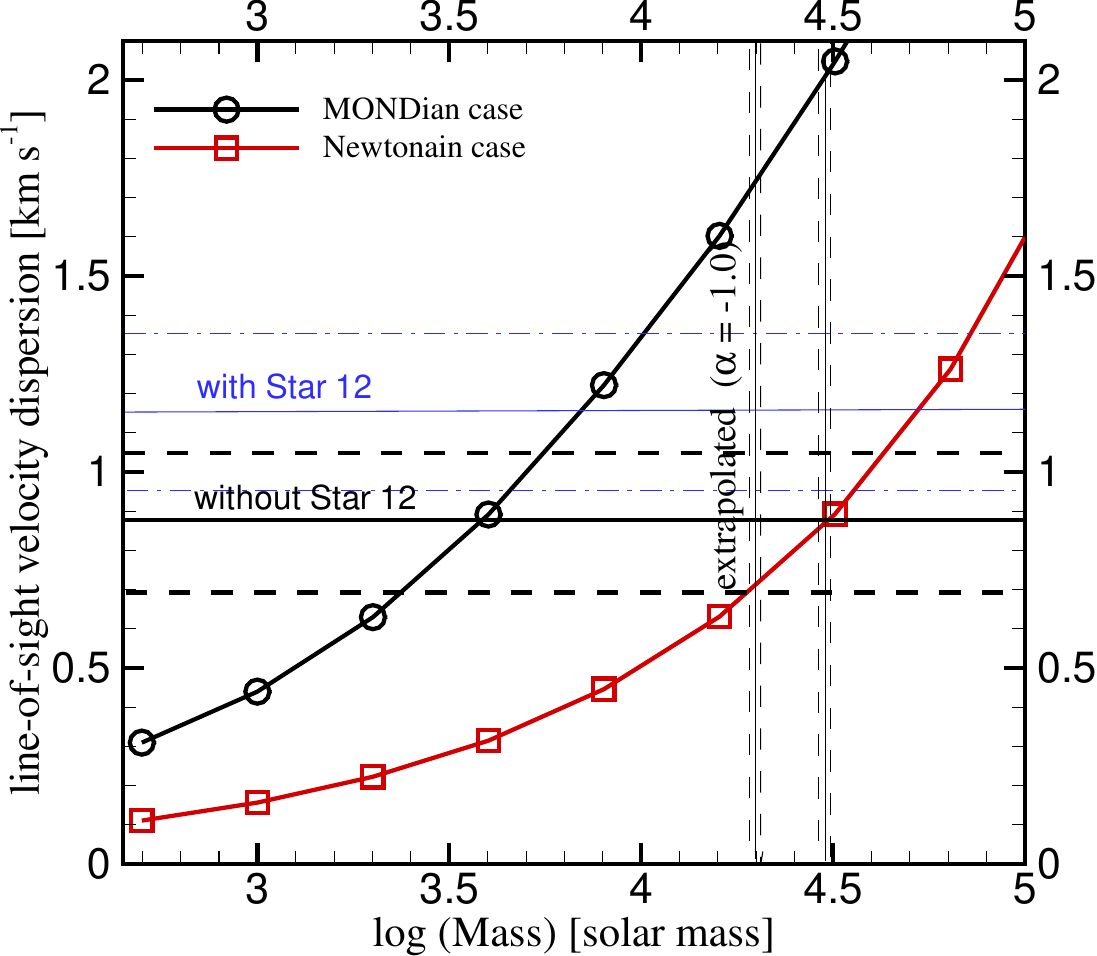}
\caption{Theoretically predicted line-of-sight velocity dispersion as a
function
of mass for the Newtonian case (red open squares) and the MONDian case (black
open
circles). 
The predictions are taken from recent $N$-body simulations by
\citet{2011A&A...527A..33H}.
The observed velocity dispersion based on the 23 clean member stars and its
uncertainty (Section~\ref{sec:specresults}) are 
shown by black solid and dashed horizontal lines, respectively. Blue solid and
dashed lines represent the velocity dispersion and uncertainty obtained when
including star\,12. For the MONDian case the predicted cluster mass when
excluding star 12 and its 1$\sigma$ uncertainty are given by
$\mathrm{M}_\mathrm{MOND}=3900^{+1400}_{-1500}\,\msun$, while in Newtonian dynamics they are
$\mathrm{M}_\mathrm{Newton}=32000\pm\,13000\msun$. Including star 12, the predicted masses
amount to $\mathrm{M}_\mathrm{MOND}=6900^{+3100}_{-2300}\,\msun$ and
$\mathrm{M}_\mathrm{Newton}=53000^{+18000}_{-16000}\,\msun$ respectively. The vertical black
lines indicate the observed total mass (solid line) and its uncertainty
(dashed lines), $\mathrm{M}_\mathrm{phot}=29800\pm800\,\msun$, and the mass derived for a
mass function significantly depleted in low-mass stars (see text),
$\mathrm{M}_\mathrm{decl,phot}=20100\pm600\,\msun$.} 
\label{fig:veldisp}
\end{figure}

\label{sec:disc}
\subsection{Newtonian and MONDian dynamical mass}
In order to see if the observed velocity dispersion and mass of Pal\,4 are
more compatible with Newtonian or MONDian dynamics, we compare the observed
global line-of-sight velocity dispersion with expected velocity dispersions
for different cluster masses for the two cases. The expected line-of-sight
velocity dispersions of Pal\,4 are taken from \citet{2011A&A...527A..33H}, who
performed $N$-body simulations
of a number of outer halo globular clusters for both Newtonian and MONDian
dynamics using the particle-mesh code \textsc{N-MODY} \citep{2009MSAIS..13...89L}.

Fig.~\ref{fig:veldisp} shows the global line-of-sight velocity dispersion as
a function of the cluster mass for the Newtonian (red open squares) and the
MONDian case (black open circles). For cluster masses below $10^5$\,M$_\odot$,
the velocity dispersion in the MONDian case is significantly larger than for
the Newtonian case since the acceleration of stars in Pal\,4 is below the
critical acceleration $a_0$ of MOND, making Pal\,4 a good test case to
discriminate between the two cases. 
For a line-of-sight velocity dispersion of $0.87\pm0.18\kms$ (shown by black
horizontal lines in the figure), obtained when excluding the probable outlier
star 12 in Section~\ref{sec:specresults}, the theoretically predicted mass in
MOND is $\mathrm{M}_\mathrm{MOND}=3900^{+1400}_{-1500}\,\msun$ and in Newtonian dynamics
$\mathrm{M}_\mathrm{Newton}=32000\pm\,13000\msun$. This corresponds to mass to light ratios of
$\mathrm{M}_\mathrm{MOND}/\mathrm{L}_{V}=0.20\pm0.08\,\msun\,\lsun^{-1}$ and 
$\mathrm{M}_\mathrm{Newton}/\mathrm{L}_{V}=1.63\pm0.67\,\msun\,\lsun^{-1}$. 
For the velocity dispersion including star 12, $\sigma=1.15\pm0.20\kms$ (shown
by blue horizontal lines in Fig.~\ref{fig:veldisp}), the theoretically
predicted mass in MOND is $\mathrm{M}_\mathrm{MOND}=6900^{+3100}_{-2300}\,\msun$
($\mathrm{M}_\mathrm{MOND}/\mathrm{L}_{V}=0.35^{+0.16}_{-0.12}\,\msun\,\lsun^{-1}$), while in
Newtonian dynamics it is $\mathrm{M}_\mathrm{Newton}=53000^{+18000}_{-16000}\,\msun$
($\mathrm{M}_\mathrm{Newton}/\mathrm{L}_{V}=2.70^{+0.93}_{-0.83}\,\msun\,\lsun^{-1}$). 

In Section~\ref{sec:photresults} we derived a cluster mass of
M$_\mathrm{phot}=29800\pm800\,\msun$ based on the photometry of Pal\,4
and assuming a Kroupa IMF for low stellar masses, and a mass of
M$_\mathrm{decl,phot}=20100\pm600\,\msun$ for the case of a declining mass
function for low-mass stars. Both values agree well with the expected value
for the Newtonian case when excluding star 12. The photometric masses are
however significantly larger than the cluster mass derived for the MONDian
case. We note, that even if the cluster did not contain any stars less massive
than $0.55\,\msun$ (or fainter than our 50 per cent completeness limit of
$\la27.6{\rm\,mag}$ in F555W), its mass of $15100\pm800\,\msun$ would
significantly exceed the MONDian prediction. 

The excellent match between photometric and (Newtonian) dynamical masses also
means that there is no need to invoke the presence of dark matter in Pal\,4,
although a small amount of dark matter cannot be excluded. As mentioned in the
introduction, Pal\,4 is similarly extended and luminous as some of our
Galaxy's ultra-faint dwarf satellites. Its M/L of $\mathrm{M}_\mathrm{Newton}/\mathrm{L}_{V}\approx
M_\mathrm{phot}/\mathrm{L}_{V}\approx1.6\,\msun\,\lsun^{-1}$, however, suggests that it
is very different from these dark matter dominated systems and a `perfectly
normal' globular cluster. This is also supported by the apparent lack of a metallicity spread in Pal\,4, whereas such a spread is detected in most dwarf satellites \citep[see the discussion in][]{2010A&A...517A..59K}.

\begin{figure}
\includegraphics[width=84mm]{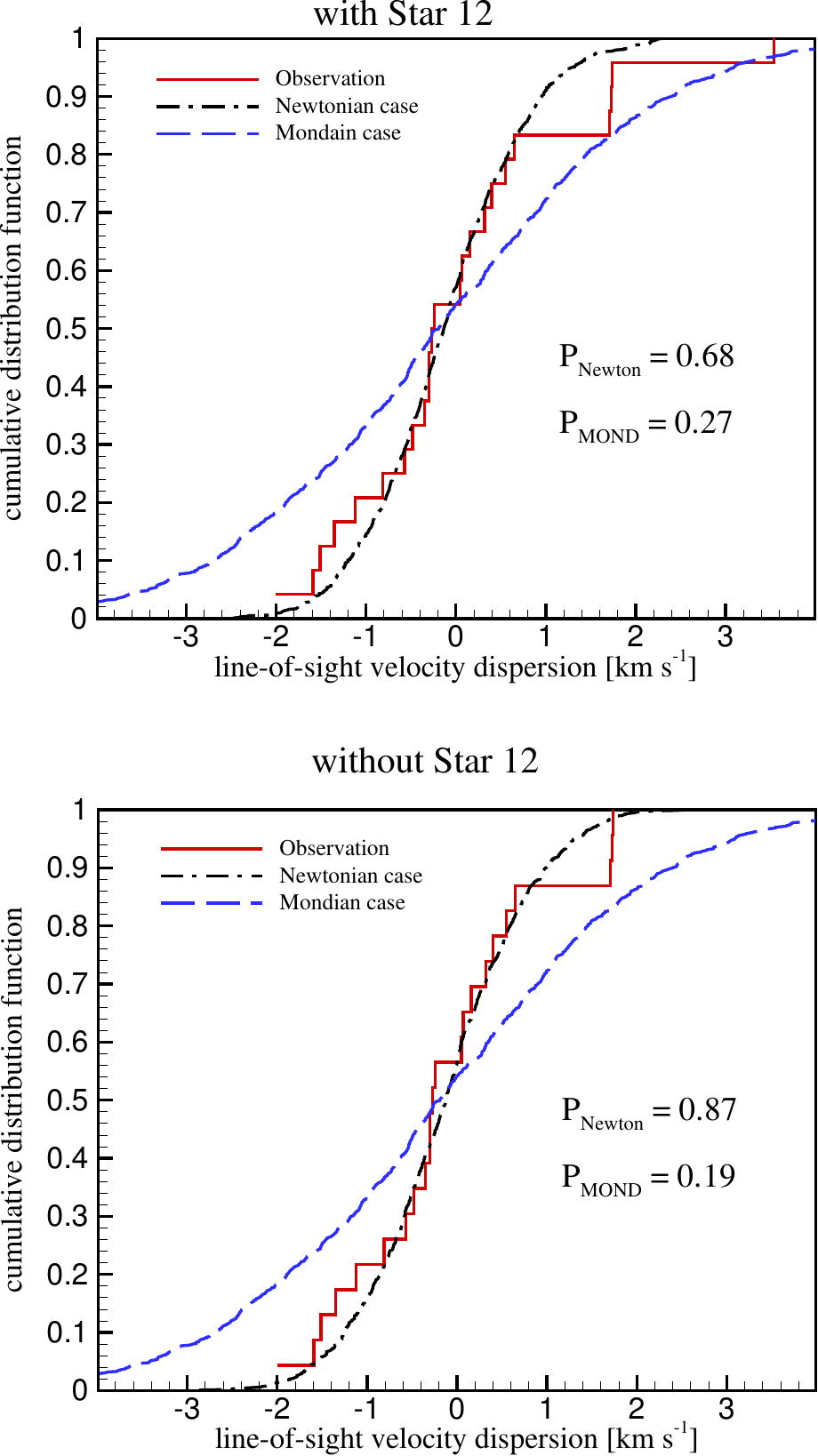}
\caption{Cumulative distribution function (cdf) of radial velocities for the
observed stars (red solid lines) and theoretical distributions
assuming Newtonian (black dashed-dotted lines) and MONDian dynamics (blue
dashed lines) and a cluster mass of $\mathrm{M}_\mathrm{phot}=29800\,\msun$. In the
upper panel, star 12 is included, in the lower panel it is excluded. The
corresponding probabilities are
shown inside the panels.}
\label{fig:ksdis}
\end{figure}

As shown by \citet{2010A&A...509A..97G}, velocity dispersions derived from a 
small sample of stars suffer from
low number statistics. We therefore used Kolmogorov-Smirnov (KS) tests to
determine the likelihood of the observed
velocity distribution in Newtonian and MONDian dynamics given the photometric
cluster mass of $\mathrm{M}_\mathrm{phot}=29800\,\msun$ 
for our sample of radial velocities either including or excluding star 12. 
Fig.~\ref{fig:ksdis} shows the resulting velocity distributions for the Newtonian
and MONDian case and the two velocity distributions. In deriving the KS
probabilities, we followed \citet{2010A&A...509A..97G} by not fixing the
systemic velocity, but shifting the model distributions in velocity such that
the maximum probability was assumed. We note that a KS test in this form is
slightly biased to favor MOND, or generally, any model predicting a higher
velocity dispersion, because it neglects the broadening of the observed
velocity distribution due to the radial velocity uncertainties. However, as
the typical velocity uncertainties in our sample are small compared to the
cluster's intrinsic velocity dispersion, the effect is small. For the
Newtonian case, a KS test gives a probability of P=0.87 if excluding star 12
and P=0.68 if including star 12. In the MONDian case, the probabilities are
P=0.19 and P=0.27 respectively. 

Apart from the stochastic effect of the small sample, our velocity dispersion
estimate is also subject to the effect of radial sampling. As two thirds of our
sample stars are located within the cluster's half-light radius, the global
velocity dispersion will be somewhat lower than our measured value. We do not
correct for this effect, but note that it will be small compared to the
statistical uncertainty because the cluster's expected velocity dispersion profile is fairly
flat (see Fig.~\ref{fig:disprad}). As a lower global velocity dispersion
will also lower the predicted masses, the discrepancy between the MONDian
prediction and the photometric mass will be larger.

The Newtonian case is therefore favored by the observational data. However,
based on the current data alone, MOND cannot be ruled out, so 
additional radial velocities will be necessary to distinguish between MONDian
and Newtonian dynamics. The simulations 
done by \citet{2011A&A...527A..33H} indicate that of order 40 radial
velocities
would be needed for Pal\,4 to 
decrease the \mbox{MONDian} P-values below 0.05 if the internal cluster
dynamics
is Newtonian. 
Nevertheless, Pal\,4 adds to the growing body of evidence that the dynamics of
star clusters in the outer Galactic halo can hardly be explained by MOND,
since the velocity dispersions of Pal\,4 (this work),
Pal\,14 \citep{2009AJ....137.4586J,2012ApJ...744..196S} and
NGC 2419 \citep{2009MNRAS.396.2051B,2011ApJ...738..186I,2011ApJ...743...43I}
are consistent with
Newtonian dynamics and below the predictions of MOND.

\begin{figure}
\includegraphics[width=84mm]{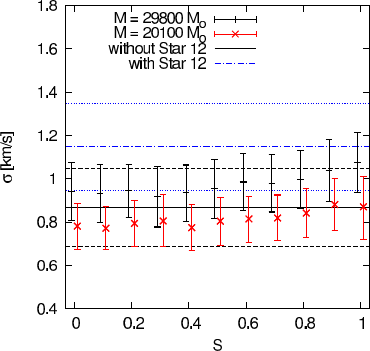}
\includegraphics[width=84mm]{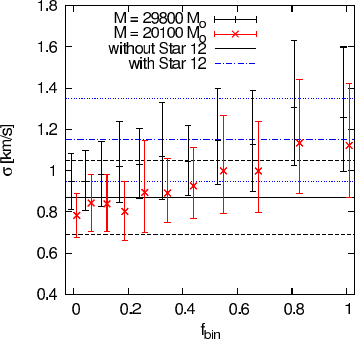}
\caption{The effect of mass segregation (upper panel) and binarity (lower
panel) on the measured velocity dispersions. Error bars show the range (68 per cent)
of velocity dispersions of samples of 23 AGB/RGB stars drawn from models of
Pal\,4. Black solid and dashed horizontal lines represent the observed velocity dispersion and its uncertainty obtained from the 23 clean member stars, blue dash-dotted and dotted lines denote the dispersion and its uncertainty derived including star\,12. Upper panel: Mass segregation can bias the measured velocity
dispersion by up to 20 per cent since AGB/RGB stars are preferentially located deeper
in the cluster potential with increasing degree of mass segregation, $S$.
Lower panel: A high binary fraction, $f_{bin}$, can severely affect the
measured velocity dispersion. Both effects may imply that Pal\,4's true
velocity dispersion is lower than the measured value, in which case the
MONDian mass estimate would be more discrepant with the observed mass.}
\label{fig:mclustersimulations}
\end{figure}

\subsection{The effect of mass segregation, unbound stars and binarity}
\label{sec:disc:effectsof}
In our analysis we did not take into account the effects of mass segregation,
of the presence of unbound stars, and of binaries.

Mass segregation will affect the interpretation of the radial velocity data in
three ways: Massive stars, such as the red giant and asymptotic giant branch
stars in our kinematic sample will reside more frequently in the cluster's
center, where the gravitational potential is deeper. Therefore they will show
a higher velocity dispersion than the global one. On the other hand, energy
equipartition will, at a given radius, cause higher mass stars to have lower
velocities, lowering the observed velocity dispersion. Moreover, in a
mass-segregated cluster, the half-mass radius is larger than the half-light
radius. Therefore, when assuming that mass follows light and equating the
half-mass radius to the observed half-light radius, a dynamical model will
overpredict the velocity dispersion. To quantify these effects, we used the
\textsc{McLuster} code \citep{2011MNRAS.417.2300K} to set up cluster models of
Pal\,4 with the characteristics obtained in this investigation. We therefore
used the best-fitting King (1966) model parameters (see
Table~\ref{tab:surfacebrightness}), a metallicity of [Fe/H]$ = -1.41$\,dex and
a cluster age of 11~Gyr. For the two photometric mass estimates, 
$\mathrm{M}_\mathrm{phot}=29800\,\msun$ and $\mathrm{M}_\mathrm{decl,phot}=20100\,\msun$, we
generated a total of 126 evolved star clusters containing a number of about
200 RGB and AGB stars each, or 130 respectively in the case of the lower mass
estimate. We set up 66 models with a varying degree of mass segregation, $S$.
We increase $S$ from zero (unsegregated) to 1.0 (completely segregated) in
steps of 0.1, where the observed degree of mass segregation in Pal\,4
corresponds approximately to a value of $0.8<S<0.9$, higher values of $S$ are
rather unrealistic. Velocity dispersions and their uncertainties were then
extracted by repeatedly drawing 23 RGB and AGB stars from the inner 100~arcsec
of the cluster models. In a similar approach as \citet{2012ApJ...744..196S}
chose for their analysis Pal\,14's velocity dispersion, we rejected stars that
differed by more than $2.5\sigma$ from the mean velocity of each sample to
emulate the clipping of likely outliers, such as star~12, in the observations.
As shown in the upper panel of Fig.~\ref{fig:mclustersimulations}, we find
that, even in the case of extreme mass segregation, the obtained velocity
dispersion rises by not more than 20 per cent compared to the non-segregated
case. The velocity dispersion we obtained for Pal\,4 may be biased by up to
10 per cent due to mass segregation. However, the error bars in
Fig.~\ref{fig:mclustersimulations} show only the 68 per cent most likely results.
Significantly higher and lower velocity dispersion measurements are still
possible with a sample of only 23 stars.

If any of the stars in the radial velocity sample are members of binary
systems, the measured velocity dispersion will be increased by the fact that
the stars are observed at a random orbital phase of the binaries. This effect
can be significant for low-mass stellar systems like Pal\,4 \citep[see,
e.g.,][]{2008A&A...480..103K,2010ApJ...722L.209M,2011ApJ...743..167B}. The
magnitude of this effect
depends on the distribution of binary periods and orbital eccentricities and
most importantly on the fraction of binaries in the cluster. 
We studied the effect of binarity by populating 60 further \textsc{McLuster}
models of Pal\,4 with a varying fraction of binaries, $f_{bin}$. We used the
same set-up as for the mass segregation models described above, but added
binaries following a Kroupa period distribution and a thermal eccentricity
distribution \citep{1995MNRAS.277.1507K}. Since periods and eccentricities
will be subject to internal dynamical evolution on a timescale of 11~Gyr, the
binaries were evolved in time with the other stars in the cluster using the
binary-star evolution routines by \citet{2002MNRAS.329..897H} that are
implemented in \textsc{McLuster}. As for the mass segregated models, velocity
dispersions were calculated from random samples of AGB and RGB stars,
rejecting velocity outliers. The results are shown in the lower panel of
Fig.~\ref{fig:mclustersimulations}. Just like mass segregation, a high
binary fraction can significantly affect the measured velocity dispersion,
resulting in a dynamical mass estimate biased towards too high masses.

Finally, unbound stars may contaminate our radial velocity sample. First of
all, energetically unbound stars, which have not yet escaped from the cluster
(so-called potential escapers) may inflate the velocity dispersion. However,
\citet{2010MNRAS.407.2241K} showed that potential escapers mainly influence
the velocity dispersion profile at large cluster radii. Moreover, also stars
within the tidal debris may be misinterpreted as bound cluster members.
\citet{2011MNRAS.413..863K} showed that for clusters in an orbital phase close
to apogalacticon the velocity dispersion may be inflated by unbound tidal
debris stars, which get pushed close to the cluster due to orbital compression
of the cluster and its tidal tails. The shallow slope of Pal\,4's surface
density profile at large cluster radii suggests that Pal\,4 may be close to
its apogalacticon, making such a contamination likely. On the other hand, this
effect may be alleviated by the fact that, because of mass segregation, the
unbound population will consist preferentially of low-mass stars, while our
radial velocity sample consists of more massive red giant and asymptotic giant
branch stars.

The combined effects of mass segregation, binaries and unbound stars render it
possible that the intrinsic velocity dispersion in Pal\,4 is lower than our
measured value of $\sigma=0.87\pm0.18\kms$. If this was the case, it would
further strengthen the case against MONDian dynamics in this cluster, as a
decreased velocity dispersion will yield an even lower cluster mass predicted
by MOND.

We note that also an anisotropic velocity distribution would affect the
velocity dispersion profile of the cluster. While the total kinetic energy is
always fixed to one half of the potential energy for a cluster in virial
equilibrium, radial anisotropy will increase the velocity dispersion in the
cluster's center compared to the isotropic case and decrease it at large
radii, and vice versa for tangential anisotropy. As our radial velocity
sample, with 15 stars inside $r_h$ and 8 stars outside $r_h$, covers a fair
range of radii, the effect of anisotropy on our measured dispersion is
expected to be only moderate. Correspondingly, \citet{2012ApJ...744..196S} in
their analysis of the similarly distributed radial velocity sample in Pal\,14,
find that the impact of even purely tangential and or maximally radial
anisotropy on the measured velocity dispersion is small.

\subsection{Primordial mass segregation}
We found clear evidence for mass segregation between main sequence
stars in Pal\,4. This mass segregation could either have evolved through
two-body relaxation and the dynamical friction of
high-mass stars or it was
already established at the time of the formation of the cluster
\citep[e.g.][]{1998ApJ...492..540H}. For a
half-light radius of 0.6 arcmin, corresponding to
18 pc, and for a cluster mass
of $\mathrm{M}=29800\,\msun$, the half-mass relaxation time of Pal\,4 is around 14~Gyr,
i.e.~of the same order as its age.
Two-body relaxation is therefore very unlikely as the explanation for the mass
segregation in Pal\,4: according to the simulations 
of \citet{2004ApJ...604..632G}, it takes several half-mass relaxation times
until a cluster with a ratio of maximum to average stellar mass of 
$\mathrm{M}_\mathrm{max}/<\mathrm{M}> \approx 4$, which is typical for a globular cluster, goes into
core
collapse. Unless Pal\,4 was significantly
 more concentrated in the past, the mass segregation in Pal\,4 was therefore
most
likely established by the star formation process itself. 

Primordial mass segregation is found in several young Galactic
\citep[e.g.][]{1988MNRAS.234..831S,1997AJ....113.1733H,2011MNRAS.413.2345H}
and Magellanic Cloud star clusters \citep[e.g.][]{1998AJ....115..592F,
2002ApJ...579..275S}. There are also indications for primordial mass
segregation in Galactic GCs: \citet{2004AJ....128.2274K} argue that the mass segregation they observed in Pal\,5 may be primordial, if the cluster that is currently being disrupted was originally a low-concentration and low-mass cluster. 
\citet{2008ApJ...685..247B} found that primordial mass segregation together
with depletion of low-mass stars by external tidal fields is necessary to
explain the present-day mass functions of stars in globular clusters.
Pal\,14, another diffuse and `young halo' cluster has a flat stellar mass
function with slope $\alpha=1.27 \pm 0.44$ within the half-light
radius \citep{2009AJ....137.4586J}, which is very similar to the slope that we
find for the center of Pal\,4. \citet{2011MNRAS.411.1989Z}
modeled the evolution of Pal\,14 over a Hubble time by direct N-body
computations on a star-by-star basis and found that in order to
reproduce its observed mass function, either strong primordial mass
segregation was necessary, or the initial mass
function (IMF) was depleted in low-mass stars. Just like in Pal\,4, the
half-mass relaxation time of Pal\,14 is comparable to its age, and
\citet{2011ApJ...737L...3B} found a non-segregated population of blue
stragglers in Pal\,14, which they interpret as observational support for the
fact that dynamical segregation has not affected the cluster yet. If one
assumes that Pal\,14 formed with a globally normal IMF, its flat central
present-day mass function found by \citet{2009AJ....137.4586J} then suggests
that the cluster had primordial mass segregation. This might hold also for
Pal\,4.
According to the simulations of \citet{2009ApJ...698..615V} long-lived
initially mass-segregated clusters should show a looser structure than
initially non-segregated clusters, as the former would lose more mass in the
central regions during early stellar evolution. It is therefore an interesting
question, if primordial mass segregation is common among diffuse GCs like
Pal\,4 and Pal\,14. 

\section{Summary}
\label{sec:concl}
We present a comprehensive analysis of the stellar mass and internal dynamics
of Pal\,4. Based on a fitting isochrones to a deep CMD and adopting literature
values for metallicity, $\alpha$-element enhancement and extinction, we
measured the cluster's age and distance to be $11\pm1$~Gyr and
$102.8\pm2.4$\,kpc respectively. Transforming stellar magnitudes to masses
using an isochrone with these parameters, we derived the cluster's mass
function from the tip of the red giant branch down to main sequence stars of
$\sim0.55\,\msun$ in the central $r<2.26$\,arcmin. The cluster shows mass
segregation, with the mass function steepening from $\alpha\la1$ inside
$r\la1.3\times r_h$ to $\alpha\ga$2.3 outside of $r\ga1.7\times r_h$. As the
cluster's half-mass relaxation time is of the order of the Hubble time, this
suggests primordial mass segregation. 

Extrapolating the measured mass function towards lower-mass stars and stellar
remnants and adopting a Kroupa mass function outside of $0.5<\mathrm{M}<1.0~\msun$, as
well as extrapolating the mass out to the cluster's tidal radius based on our
surface density profile (Section~\ref{sec:surfacebrightness}), we obtain a
total stellar mass of $\mathrm{M}_\mathrm{phot}=29800\pm800\,\msun$.

This is in excellent agreement with the dynamical mass obtained with Newtonian
dynamics, $\mathrm{M}_\mathrm{Newton}=32000\pm\,13000\msun$, based on the cluster's observed
velocity dispersion of $0.87 \pm 0.18\kms$ derived from radial velocities of
23 clean member stars. The dynamical mass predicted by MOND,
$\mathrm{M}_\mathrm{MOND}=3900^{+1400}_{-1500}\,\msun$, is significantly below the observed
stellar mass. However, in a KS test comparing the observed distribution of
radial velocities with that predicted in MONDian dynamics, MOND is also
compatible with the data at a probability of $20$ per cent. 

Thus the observational data favor Newtonian dynamics, but an extended sample
of radial velocities is needed to confidently rule out MOND, if the cluster is
governed by Newtonian dynamics.

\section*{Acknowledgments}
We thank Katrin Jordi for providing her SDSS-based surface density profile of
Pal\,4. We also thank the referee of this paper, Antonio Sollima, for the
useful report that helped improve the manuscript.

This work was partially supported by Sonderforschungsbereich 881, ``The Milky
Way System'' (subprojects A2 and A3) of the German Research Foundation (DFG)
at
the University of Heidelberg.

H.B. acknowledges support from the Australian Research Council through Future
Fellowship grant FT0991052.

S.G.D. acknowledges a partial support from the NSF grants AST-0407448 and
AST-0909182.

Some of the data presented herein were obtained at the W.M. Keck Observatory,
which is operated as a scientific partnership among the California Institute
of
Technology, the University of California and NASA. The Observatory was made
possible by the generous financial support of the W.M. Keck Foundation. 

Based in part on observations made with the NASA/ESA Hubble Space Telescope, obtained
from the Multimission Archive at the Space Telescope Science Institute (MAST).
STScI is operated by the Association of Universities for Research in
Astronomy,
Inc., under NASA contract NAS5-26555. 

This research has made use of the NASA/ IPAC Infrared Science Archive, which
is
operated by the Jet Propulsion Laboratory, California Institute of Technology,
under contract with NASA.

\bibliographystyle{mn2e}
\bibliography{pal4phot}

\bsp

\label{lastpage}

\end{document}